\documentclass[twocolumn,aps,prx,floatfix,longbibliography,groupedaddress,nofootinbib]{revtex4-1}
\usepackage{hyperref}
\usepackage{amsfonts}
\usepackage{amsbsy}
\usepackage{xcolor}
\usepackage{soul}
\usepackage{graphicx}
\usepackage{mathtools}

\newcommand{\bra}[1]{\ensuremath{\left\langle #1 \right\vert}}
\newcommand{\ket}[1]{\ensuremath{\left\vert #1 \right\rangle}}

\newcommand{\spvec}[1]{\ensuremath{\mathbf{#1}}}

\newcommand{\unitvec}[1]{\hat{\mathbf{{#1}}}}

\newcommand{\bea}{\begin{eqnarray}}
\newcommand{\eea}{\end{eqnarray}}
\newcommand{\beq}{\begin{equation}}
\newcommand{\eeq}{\end{equation}}

\newcommand{\rv}{{\bf r}}

\newcommand{\Dc}{{\cal D}}

\newcommand{\Pc}{\mathcal{P}}

\newcommand{\commentout}[1]{{}}
\newcommand{\radKernel}{\ensuremath{\mathsf{G}}}

\usepackage{float}

\begin{document}
\title{Subradiance-protected excitation spreading in the generation of collimated photon emission from an atomic array}
\date{\today}
\author{K.~E.~Ballantine}
\email{k.ballantine@lancaster.ac.uk}
\author{J.~Ruostekoski}

\email{j.ruostekoski@lancaster.ac.uk}

\affiliation{Department of Physics, Lancaster University, Lancaster, LA1 4YB, United Kingdom}

\begin{abstract}

We show how an initial localized radiative excitation in a two-dimensional array of cold atoms can be converted into highly-directional coherent emission of light by protecting the spreading of the excitation across the array in a subradiant collective eigenmode with a lifetime orders of magnitude longer than that of an isolated atom. We demonstrate how to reach two such strongly subradiant modes, a uniform one where all the dipoles are oscillating in phase normal to the plane and an antiferromagnetic mode where each dipole is $\pi$ out of phase with its nearest neighbor.
The excitation, which can consist of a single photon, is then released from the protected subradiant eigenmode by controlling the Zeeman level shifts of the atoms. Hence, an original localized excitation which emits in all directions is transferred to a delocalized subradiance-protected excitation, with a probabilistic emission of a photon only along the axis perpendicular to the plane of the atoms. This protected spreading and directional emission could potentially be used to link stages in a quantum information or quantum computing architecture.
\end{abstract}

\maketitle

\section{Introduction}

Collective optical interactions in cold trapped atomic ensembles, where systems with increasing densities can now be achieved, are being actively studied experimentally~\cite{BalikEtAl2013,CHA14,Pellegrino2014a,Havey_jmo14,wilkowski,
Jennewein_trans,wilkowski2,Ye2016,Jenkins_thermshift,
vdStraten16,Guerin_subr16,Machluf2018,Dalibard_slab,Bettles18}. Since the light-mediated resonant dipole-dipole interactions depend on the average atomic separation, the close proximity of the atoms can lead to a correlated optical response~\cite{Morice1995a,Ruostekoski1997a} that no longer obeys continuous medium descriptions of electrodynamics~\cite{Javanainen2014a,JavanainenMFT}. The quest for systems where the collective optical response could potentially be easily manipulated has resulted in the studies of strong light-atom coupling -- and coupling of light with other dipolar resonators -- in regular arrays~\cite{lemoult2010,Fedotov2010,Jenkins2012a,Perczel,Bettles_17topo,CAIT,Bettles2015d,Bettles2016,Facchinetti16,valentine2014,Facchinetti18,Shahmoon,Plankensteiner2017,Asenjo-Garcia2017a,Jen17,Guimond2019,Ritsch_subr,Grankin18,Olmos13,Kramer2016,Sutherland1D,Yoo2016,Wang2017,Jenkins17,Yoo18,Imamoglu2018b,Manzoni_2018,Mkhitaryan2018,Jen18,Zhang2018,Henriet2018,Parmee2018,Plankensteiner19,Javanainen19}.

Control, storage and transmission of collective excitations could play a key role in modular quantum architecture~\cite{ChoiEtAlNature2008,Duan10,Nickerson14}, consisting of individual quantum elements with coherent links~\cite{Grankin18,Monroe14}. Subradiant states~\cite{Dicke54}, which decay more slowly than an isolated emitter, couple weakly to external fields and have
posed a long-standing experimental challenge, with observations first emerging in two- and few-particle systems~\cite{DeVoe,Hettich,McGuyer,Takasu,Lovera,Frimmer}, and now also in larger ensembles~\cite{Guerin_subr16,Jenkins17}. Due to the isolation from the environment, subradiant modes have been shown to be useful in transport of excitations~\cite{Willingham11,Giusteri15,Leggio_2015,Doyeux17}. Recent work has explored light transport for closely spaced atoms in arrays with topological edge states~\cite{Perczel,Bettles_17topo} and in a one-dimensional (1D) chain or ring~\cite{Jen_2019,Cardoner19,Needham19}.

We show here how a strongly protected spreading of a radiative excitation can be achieved in a two-dimensional (2D) planar array of cold atoms. The protocol involves manipulating the internal atomic level shifts to produce an initial excitation having a sufficient overlap with extremely subradiant collective excitation eigenmodes. The initial localized excitation at the center of the array, which will emit radiation in all directions, is then transported across the lattice and converted into highly directional emission. The initial localized excitation can be generated by a single-photon source and will generally have an overlap with several of the collective radiative many-atom excitation eigenmodes of the system. These collective modes arise from the effective dipole-dipole interactions between atoms, and will each have a collective resonance shift and a collective linewidth that differs from that of a single atom. As these eigenmodes have different linewidths, they will decay at different rates, and interference of the modes leads to the spatial spreading of the excitation across the lattice. After some time, the most subradiant eigenmode of the initial single-photon excitation will become dominant, and will be delocalized over the entire array. In this case coherence across the array comes from the collective nature of the many-atom eigenmode, rather than any driving field. After we release the excitation by controlling the atomic levels, the result is coherent and highly directional emission of a photon from the array.

In our model, we assume that a 2D square array of cold atoms is prepared with a single atom per lattice site and with a single-photon excitation initially localized at the center of the lattice. Such excitations could be produced by coupling to a nearby emitter or by selectively controlling the atomic excitation. By choosing the initial state and the lattice spacing appropriately, this local excitation is transferred to one of two delocalized subradiant target modes, a mode with uniform in-phase out-of-plane polarization, and a mode where the phase of the out-of-plane polarization varies by $\pi$ between neighboring atoms,  an antiferromagnetic (AFM) mode. Once the delocalized mode is established, Zeeman splitting can couple either of the subradiant out-of-plane mode to a uniform in-plane mode that is strongly radiating, allowing the excitation to decay. We calculate the far-field radiation pattern of this emission which is highly collimated in the direction normal to the plane. Collimated emission has been achieved in plasmonic planar arrays after spreading of an initial excitation~\cite{AdamoEtAlPRL2012}, but without an analogous procedure of transferring the excitation between weakly (dark) and strongly (bright) radiating states. An off-resonant pair of atomic layers, where the atoms do not experience strong resonant dipole-dipole interactions, has been proposed as a phased-array antenna for single photons~\cite{Grankin18}.


\section{Atom-Light coupling}
\label{sec:coupling}

\subsection{Single excitation model}

\begin{figure}
\centering
		\includegraphics[width=1\columnwidth]{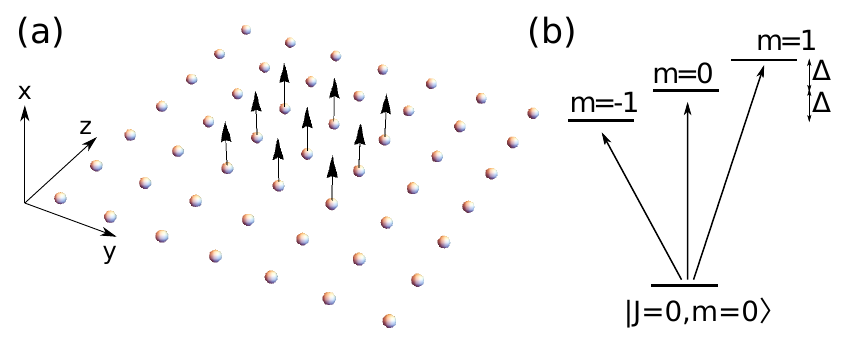}
    \vspace*{-12pt}
		\caption{
		(a) Illustration of planar array of atoms with an initial single-photon excitation localized on a small number of atoms at the center of the array. The arrows represent the direction of the atomic dipoles that are oriented normal to the plane and will propagate toward a uniform subradiant eigenmode. (b) The atomic level structure where the Zeeman splitting $\Delta$ is only introduced for the release of the subradiant excitation.
		}
		\label{fig:intro}
\end{figure}

We consider a 2D square array of $N$ atoms in the $y z$ plane, with the lattice spacing $d$, and a $J=0\rightarrow J'=1$ atomic transition (Fig.~\ref{fig:intro}).
The full quantum dynamics of the atomic system for a given initial excitation and in the absence of a driving laser follows from the quantum master equation
 for the many-atom density matrix $\rho$~\cite{Lehmberg1970}:
 \begin{equation}
\begin{multlined}
\label{eq:rhoeom}
\dot{\rho} = i\sum_{j,\nu}\Delta_{\nu}\left[\hat{\sigma}_{j\nu}^{+}\hat{\sigma}_{j\nu}^{-},\rho\right] +i\sum_{jl\nu\mu (l\neq j)}\Omega^{(jl)}_{\nu\mu}\left[\hat{\sigma}_{j\nu}^{+}\hat{\sigma}_{l\mu}^{-},\rho\right] \\
+\sum_{jl\nu\mu}\gamma^{(jl)}_{\nu\mu}\left(
2\hat{\sigma}^{-}_{l\mu}\rho\hat{\sigma}^{+}_{j\nu}-\hat{\sigma}_{j\nu}^{+}\hat{\sigma}_{l\mu}^{-}\rho -\rho\hat{\sigma}_{j\nu}^{+}\hat{\sigma}_{l\mu}^{-}\right) \,,
\end{multlined}
\end{equation}
where $\hat{\sigma}_{j\nu}^{+}=(\hat{\sigma}_{j\nu}^{-})^\dagger=\ket{e_{j\nu}}\bra{g_j}$ is the raising operator to the excited state $\nu$ on atom $j$ \footnote{The atom and light field amplitudes in this paper refer to slowly varying amplitudes, where the rapid oscillations at the resonance frequency have been factored out, e.g., $\exp(-i \omega t)$ from $\hat{\sigma}_{j\nu}^{-}$.}. The single atom Zeeman splittings, $\Delta_\nu$, are given relative to the $m=0$ transition frequency $\omega$, with $\Delta_0=0$ and $\Delta_{-}=-\Delta_{+}\equiv\Delta=\omega-\omega_{\pm}$, where $\omega_{\pm}$ are the frequencies of the $m=\pm1$ transitions, respectively. The diagonal terms of the dissipative matrix $\gamma^{(jj)}_{\nu\nu}=\gamma=\mathcal{D}^2k^3/(6\pi\hbar\epsilon_0)$ correspond to the single atom resonance linewidth with the reduced dipole matrix element $\mathcal{D}$ and $k=2\pi/\lambda=\omega/c$. The off-diagonal elements in the dissipation and interaction terms are given by the real and imaginary parts of
\begin{equation}
\Omega^{(jl)}_{\nu\mu}+i\gamma^{(jl)}_{\nu\mu}=
\xi {\cal G}^{(jl)}_{\nu\mu} \, ,
\end{equation}
where $\xi=6\pi\gamma/k^3$. Here, the dipole-dipole interaction between the atoms $j$ and $l$ with the orientations of the dipoles $\unitvec{e}_{\nu}$ and $\unitvec{e}_{\mu}$  at positions $\rv_j$ and $\rv_l$, respectively,  is determined by the dipole radiation kernel~\cite{Jackson}
\begin{equation}
  \label{eq:V_2_level_def}
  \mathcal{G}^{(jl)}_{\nu\mu} = \unitvec{e}^\ast_{\nu} \cdot
  \radKernel'(\rv_j-\rv_l) \unitvec{e}_{\mu} \textrm{,}
\end{equation}
where the familiar expression for the positive frequency component of the monochromatic dipole radiation reads
\begin{equation}
  \label{eq:rad_kernel_def}
  \radKernel_{\alpha\beta}(\spvec{r}) = \left[\frac{\partial}{\partial r_\alpha}
    \frac{\partial}{\partial r_\beta} - \delta_{\alpha\beta}\nabla^2\right]
  \frac{e^{ikr}}{4\pi r} -\delta_{\alpha\beta}\delta(\rv) \textrm{ ,}
\end{equation}
and the contact interaction term between the atoms has explicitly been removed~\cite{Lee16}
\begin{equation}
\radKernel_{\alpha\beta} ^\prime(\rv)= \radKernel_{\alpha\beta} (\rv)+\frac{\delta_{\alpha\beta}\delta(\rv)}{3}.
\end{equation}
For ideal point dipoles this contact interaction is inconsequential for the physics~\cite{Ruostekoski1997b}, and by assuming hard-core atoms it vanishes as the atoms cannot overlap. We assume that the Zeeman level splitting $\Delta$ of the $J'=1$, $m=\pm 1$ states, is controllable, as could be obtained, e.g.,  by AC-Stark shifts of lasers or microwaves~\cite{gerbier_pra_2006} or by magnetic fields. For the majority of what follows, we take $\Delta=0$ and study the transfer and decay of an initial excitation in the regime where any Zeeman splitting is negligible. In Sec.~\ref{ss:single_atom} we describe the effects of Zeeman splitting on a single atom. This is relevant in Sec.~\ref{ss:spreading}, where we consider how this splitting can be used to create the initial excitation, and in Sec.~\ref{sub:release} where we show how Zeeman splitting couples a subradiant state, with very slow decay, to a rapidly decaying bright mode, allowing the photon to radiate away.

Instead of solving the general quantum master equation for the full multi-excitation dynamics, we restrict ourselves to single-excitation systems determined by the initial state of precisely one electronic excitation and study their evolution. Such a localized  excitation can be achieved by short-range coupling to a single qubit in the excited state, as discussed in Sec.~\ref{ss:spreading}.
The dynamics of the single-excitation subspace decouples to give 
\begin{equation}
\partial_t\bar{\rho}^{(jk)}_{\nu\mu} = i\mathcal{H}_{\nu\tau}^{(jl)}\bar{\rho}^{(lk)}_{\tau\mu}-i\bar{\rho}^{(jl)}_{\nu\tau}\mathcal{H}_{\tau\mu}^{(lk)\ast},
\end{equation} 
where $\bar{\rho}^{(jk)}_{\nu\mu}=\bra{G}\hat{\sigma}^{-}_{j\nu}\rho\hat{\sigma}^{+}_{k\mu}\ket{G}$ are the matrix elements of $\rho$ corresponding to the single excitation, $\ket{G}$ is the state with all atoms in the ground state, and
\begin{equation}
\mathcal{H}^{(jk)}_{\mu\nu} = \Delta_\mu\delta_{jk}\delta_{\mu\nu}+\Omega^{(jk)}_{\mu\nu}(1-\delta_{jk})+i\gamma^{(jk)}_{\mu\nu}.
\end{equation}

While dissipation will mix the single-excitation subspace with the ground state, the dynamics within the single-excitation subspace are coherent. Since we always assume a pure initial state, the single-excitation part of the density matrix retains the form 
\begin{equation}
\bar{\rho}(t) = \ket{\Psi(t)}\bra{\Psi(t)} ,
\end{equation}
where $\ket{\Psi(t)}$ is a single-excitation state whose norm $|\langle \Psi(t) |\Psi(t)\rangle|^2$ is not conserved due to the dissipation. This state can be expanded in terms of the atomic excitations
\begin{equation}
\ket{\Psi(t)} = \sum_{j,\nu} \Pc^{(j)}_{\nu}(t)\,\hat{\sigma}^{+}_{j\nu} \ket{G}.
\end{equation}
Regarding single-particle expectation values, equations of motion can equivalently be written in terms of these single-excitation subspace amplitudes~\cite{SVI10}. Here we express this by rearranging the excitation amplitudes into a vector ${\bf b}_{3j-1+\nu}=\Pc^{(j)}_{\nu}$  ($\nu=-1,0,1$), and similarly arranging the evolution Hamiltonian into a $3N\times3N$ matrix $\mathcal{H}^\prime_{3j-1+\nu,3k-1+\mu}=\mathcal{H}^{(jk)}_{\nu\mu}$. Then the  time evolution is described by
\beq
\dot{{\bf b}} = i \mathcal{H}^\prime{\bf b} ,
\label{eq:eigensystem}
\eeq
where $\mathcal{H}^\prime$ is a non-Hermitian matrix.

This matrix equation describing the dynamics of the amplitudes of a single excitation state is formally equivalent to the equations of motion of $N$ classical linear coupled dipoles. Thus we can interpret the results in terms of the decay of a single photon, or the decay of a classical coherent dipolar excitation. For a dipole  $\Pc^{(j)}_{\nu}$ of the $j$th atom, the scattered field at $\rv$ reads~\cite{Jackson}
\begin{equation}
  \label{eq:e_field}
  \epsilon_0\spvec{E}^{(j)} (\rv)= \radKernel(\rv - \rv_j)\,\Dc\sum_{\nu} \unitvec{e}_{\nu}\Pc^{(j)}_{\nu}\textrm{.}
\end{equation}

In order to understand the behavior of the system as it evolves away from the initial condition, we study the eigenvalues $\delta_n+i \upsilon_n$  and eigenmodes ${\bf v}_n$ of the coupling matrix $\mathcal{H}^\prime$,
where $\delta_n=\omega-\omega_n$ is the shift of the collective mode resonance from that of a single atom and $\upsilon_n$ is the collective resonance linewidth.

Since $\mathcal{H}^\prime$ is not Hermitian, the eigenmodes will not be orthogonal. However, we find that in all our numerical simulations they still form a complete basis. One can therefore uniquely express
the excitation amplitudes as
\begin{equation}
{\bf b} (t) = \sum_{n} c_n(t) {\bf v}_n.
\label{EXPANSION}
\end{equation}
The amplitudes {$c_{n}$} of the collective modes evolve independently and each satisfy the equation of motion
\begin{equation}
  \label{eq:eqOfMEignModes}
  \dot{c}_n = \left(i\delta_n - \upsilon_n \right) c_n ,
\end{equation}
 with solution 
\begin{equation}
\label{eq:EignModeSol}
c_n(t) = \exp{\left[t (i\delta_n-\upsilon_n)\right]} c_n(0).
\end{equation}
Due to the non-orthogonality of the eigenvectors, we use the definition
\beq
L_j(t)= {|{\bf v}_j^T {\bf b(t)}|^2\over \sum_i | {\bf v}_i^T  {\bf b(0)} |^2}\,,
\label{eq:measure}
\eeq
for a measure of the occupation of an eigenmode ${\bf v}_j$ in the current state ${\bf b}$. This has been shown to describe accurately the contribution of the collective mode occupations in the decay dynamics of radiative excitations~\cite{Facchinetti16}.  This measure can also be used to determine a target eigenmode of a finite lattice which most closely matches an ideal uniform mode with equal amplitude on every site.

\subsection{Effects of Zeeman splitting}
\label{ss:single_atom}

In the absence of Zeeman splitting the $J=0\rightarrow J^\prime=1$ transition is isotropic. This symmetry is broken by the plane of the atomic lattice, and collective modes can be grouped into those where the polarization is out-of-plane and those where it is in plane. Zeeman splitting rotates the polarizations around an effective magnetic field transferring the excitation between in-plane and out-of plane modes~\cite{Facchinetti16}. We use this effect in Sec.~\ref{ss:spreading} to create an initial localized out-of-plane excitation, and in Sec.~\ref{sub:release} to couple the delocalized out-of-plane-excitation to an in-plane excitation, allowing the light to radiate away.

We illustrate the effect of Zeeman splitting (along the $z$ direction)  in coupling the $x$ and $y$ polarizations by first considering a single isolated atom $j$. The transitions $\ket{J=0,m=0}\rightarrow\ket{J^\prime=1,m=\pm 1}$ correspond to the unit circular polarization vectors $\unitvec{e}_{\pm}$. For a single atom the polarizations $\mathcal{P}_{\pm 1,0}^{(j)}$ evolve independently. However, if we instead write the equations of motion in the Cartesian basis;
\begin{align}
\mathcal{P}_{x}^{(j)} &= \frac{1}{\sqrt{2}}\left(\mathcal{P}_{-1}^{(j)}-\mathcal{P}_{+1}^{(j)}\right), \\
\mathcal{P}_{y}^{(j)} &= -\frac{i}{\sqrt{2}}\left(\mathcal{P}_{-1}^{(j)}+\mathcal{P}_{+1}^{(j)}\right),
\end{align}
then we obtain~\cite{Facchinetti16,Facchinetti18}
\begin{align}
\label{zeeman1}
\partial_t \mathcal{P}_{x}^{(j)} &= (i\Delta_0-i\delta_s-\gamma)\mathcal{P}_{x}^{(j)} - \delta_a\mathcal{P}_{y}^{(j)}, \\
\label{zeeman2}
\partial_t \mathcal{P}_{y}^{(j)} &= (i\Delta_0-i\delta_s-\gamma)\mathcal{P}_{y}^{(j)} + \delta_a\mathcal{P}_{x}^{(j)},
\end{align}
where $\delta_s=-(\Delta_{+}+\Delta_{-})/2=0$ and $\delta_a=(\Delta_{-}-\Delta_{+})/2=\Delta$ for the case of symmetric splitting. 
Hence, the two amplitudes are coupled and an initial polarization in the $y$ direction will act as a source of $x$ polarization, and vice-versa. In the absence of Zeeman splitting the isotropy of the $J=0\rightarrow J^\prime=1$ means that any orientation of the basis vectors forms an eigenbasis. However, for $\Delta\neq 0$ the isotropy is broken; $\mathcal{P}_{x,y}$ are no longer eigenstates and will rotate around the effective magnetic field generating the Zeeman shift. In practise, the shift can be faster to control using microwave- or laser-induced AC-Stark shifts~\cite{gerbier_pra_2006}) than magnetic fields. The sign of $\Delta$ determines the direction of this rotation. Hence, depending on this sign, a component $\Pc_y$ can be rotated toward either $\pm\Pc_x$, which represent the same dipole orientation but with a $\pi$ phase difference. By preparing different atoms in different internal states the sign of the Zeeman splitting, and hence the orientation of the coupling between $\Pc_y$ and $\Pc_x$, can be chosen to vary from site to site. In this way uniform polarization in the $y$ direction on each atom can be coupled to two different out-of-plane modes which will be discussed further in Sec.~\ref{ss:modes}, a uniform mode where each dipole points in the $x$ direction with the same phase, and an AFM one where each dipole points in the $x$ direction with the phase varying by $\pi$ between each atom and its nearest neighbors.

\subsection{Initialization of localized excitation}

\label{ss:initial}

In the analysis of the subradiance-protected excitation spreading and collimated photon release, we first study the initialization of a localized excitation which, for simplicity, we take extend over nine sites.
We will consider two types of initial excitation, one where the central nine atoms begin with polarization in the $x$ direction which is in phase, and one where this initial polarization varies by a phase of $\pi$ between neighboring atoms. To model the initialization of these localized lattice excitations by a single-photon source, we add the amplitude of the qubit excitation $\mathcal{Q}$ to the dynamics described by~Eq.~(\ref{eq:eigensystem}), 
\begin{align}
\partial_t\mathcal{Q} &= -iJ_{j\mu}\mathcal{P}_{\mu}^{(j)}, \\
\partial_t\mathcal{P}_{\mu}^{(j)} &= -iJ_{j\mu}^\ast\mathcal{Q} + i\mathcal{H}_{\mu\nu}^{(jk)}\mathcal{P}_{\nu}^{(k)}, 
\end{align}
where $J_{j\mu}$ is the coupling of the qubit to the polarization $\mu$ on atom $j$. Here we use a coupling that is approximately constant over a few atoms at the center of the lattice, i.e., over distances of the order of a few wavelengths. One example of a possible implementation is using  a laser-assisted coupling and highly excited Rydberg states. As in similar proposals~\cite{Petrosyan18,Grankin18}, the dressing laser provides off-resonance coupling between the ground state of the lattice atoms and a Rydberg state, which then has a dipole-dipole coupling to Rydberg states of the qubit atom. In that case the coupling 
\begin{equation}
J_{j\mu} = J_0\Omega_d(\rv_j) [\unitvec{e}_\mu^\ast\cdot \radKernel'(\rv_j-\rv_q){\bf d}_q]
\end{equation}  
is proportional to the dipole-dipole interaction with the qubit dipole ${\bf d}_q$ at position $\rv_q$, and the spatially dependent Rabi frequency of the dressing laser $\Omega_d(\rv_j)$.

We take the qubit to be displaced from the lattice in the $x$ direction, and have a dipole moment in the $y$ direction. This primarily drives polarization in the $y$ direction in the lattice atoms, with equal phase. To initialize a localized excitation with polarization in the $x$ direction, Zeeman splitting is used to rotate the polarization, as discussed in Sec.~\ref{ss:single_atom}. If the direction of the rotation is the same for all atoms, this initial state will then have in-phase polarization in the $x$ direction. Alternatively, by varying the sign of the Zeeman splitting $\Delta$, the $y$ polarization can be made to rotate in opposite directions on different sites.
The spatial variation of the level shift could possibly be created by directly adjusting the atom traps or by the AC Stark shift of an external standing-wave laser with the intensity varying along the diagonal of the array (along the direction $\hat{\bf y}+ \hat{\bf z}$) with the intensity maxima separated by $\sqrt{2} d$.\footnote{Suitable transitions could be found, e.g., with Sr or Yb. For the $^3P_0\rightarrow ^3D_1$ transition of $^{88}$Sr~\cite{Olmos13}, the resonance wavelength $\lambda\simeq 2.6\mu$m and the linewidth $2.9\times 10^{5}$/s. For the case of optical lattices the array spacing $d$ at magic wavelength for the same transition can be as short as 206.4nm, yielding $d/\lambda\simeq 0.08$. The optical lattice spacing for a 2D planar lattice can also be controlled to achieve the right periodicity by tilting the propagation direction of the lasers forming the lattice.}
This results in an initial localized excitation with the checkerboard-pattern of the phase of $\Pc_x$ differing by $\pi$ between different sites. 
An alternative approach is to have atoms in different hyperfine states occupying the different lattice sites~\cite{mandel03}, with the associated description of the atom-light dynamics~\cite{Jenkins_long16,Lee16}.  Independently of which eigenmode we are targeting, a simple numerical example shows that a single photon from the nearby qubit can in principle be transferred accurately very close to the desired initial state. 

\subsection{Excitation spreading}

\label{ss:spreading}

We now study numerically the evolution and decay of an initial localized excitation consisting of a single photon. We examine how the appropriately prepared excitation spreads across the lattice due to a subradiant collective mode and comes to occupy a target delocalized mode.

\subsubsection{Collective modes}
\label{ss:modes}
Once the initial excitation (Sec.~\ref{ss:initial}) is established, the Zeeman shift is turned off, and the state evolves with $\Delta=0$ until we consider releasing the excitation in Sec.~\ref{sub:release}. While the excitation decays to zero in the long-time limit, it is clear from Eq.~(\ref{eq:EignModeSol}) that if there is appreciable initial excitation of a subradiant state with $\upsilon_n\ll \gamma$, then this will dominate at intermediate times when other mode amplitudes have decayed. There are two such modes that we target here. The first consists of coherent uniform in-phase polarization which points perpendicular to the plane, along the $x$ axis, and which we denote by $\Pc_P$. This eigenmode can be targeted with an appropriate initial excitation of in-phase $x$ polarization, even when this initial excitation is highly localized. Here we also target another collective eigenmode, an AFM excitation denoted by $\Pc_{AF}$, where each atom has a polarization in the normal direction along the $x$  axis which is $\pi$ out of phase with each of its nearest neighbors. This mode is of interest because it has quickly varying phase across the lattice, but can be reached from a localized initial excitation where the polarization of almost all the atoms is zero. This localized excitation consists of alternating polarization $\Pc_x^{(j)}=\pm 1/3$ on the central nine atoms. 

We also consider a third mode, denoted $\Pc_I$, which has a uniform in-phase polarization in-plane, here chosen to be along the $y$ axis. The eigenmode $\Pc_I$ directly couples to an incident plane wave propagating normal to the plane and is generally very easy to excite. Similarly to the single atom case, the out-of-plane eigenmodes $\Pc_P$ and $\Pc_\mathrm{AF}$ can be directly coupled to $\Pc_I$ by a symmetry breaking where induced Zeeman level shifts drive transitions between the modes~\cite{Facchinetti16,Facchinetti18}. The uniform modes have previously been shown to be useful in understanding the spectral response of the array, and can play the role of collective versions of dark and bright states~\cite{Facchinetti18} of a standard single-particle electromagnetic induced transparency (EIT)~\cite{FleischhauerEtAlRMP2005}.

\begin{figure}
\centering
		\includegraphics[width=1\columnwidth]{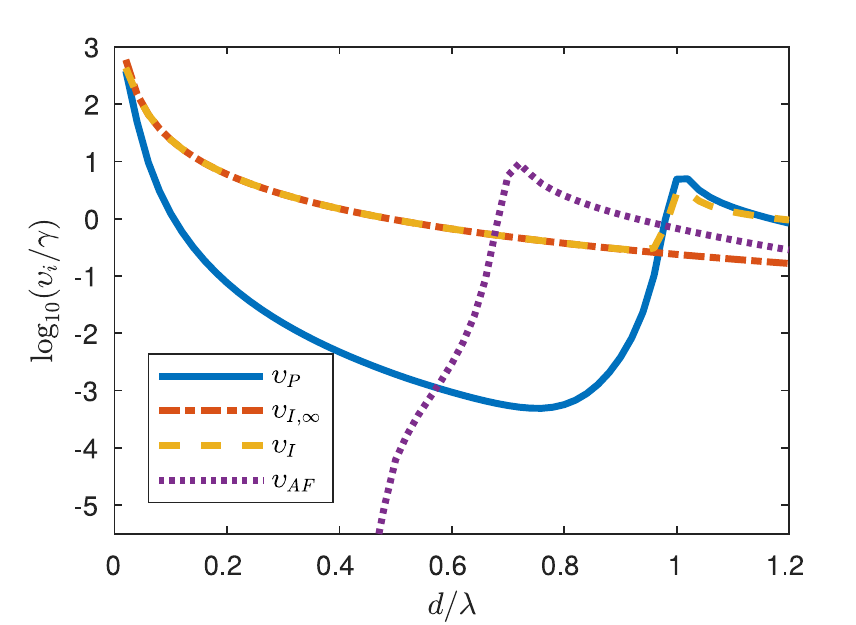}
    \vspace*{-12pt}
		\caption{
		Collective linewidth as a function of lattice spacing of the uniform out-of-plane eigenmode ($\upsilon_P$), the uniform-in-plane eigenmode ($\upsilon_I$), and the antiferromagnetic (AFM) out-of-plane eigenmode ($\upsilon_\mathrm{AF}$) with the phase of polarization varying by $\pi$ between nearest neighbors. The linewidth of the AFM mode falls rapidly when it crosses the light line at $d=\lambda/\sqrt{2}$, where it would go to zero in the $N\rightarrow \infty$ limit. The dashed-dotted line shows the analytic infinite-lattice limit formula for $\upsilon_I$ given by Eq.~(\ref{eq:inplanelinewidth}), which is valid only for $d<\lambda$. 
		}
		\label{fig:linewidthvspacing}
\end{figure}

The resonance linewidth of each of these modes is shown in Fig.~\ref{fig:linewidthvspacing} as a function of lattice spacing for a $31\times 31$ lattice. The uniform out-of-plane mode $\Pc_P$ and the AFM mode $\Pc_{AF}$ can be strongly subradiant at the appropriate array spacings. The AFM mode has maximum phase variation in the $y$ and $z$ directions, and so for periodic boundary conditions is located at the corner of the Brillouin zone with quasi-momentum ${\bf q}=(\pi/d,\pi/d)$. For small lattice spacing the quasi-momentum here is greater than the free-space wavevector, $|{\bf q}|>k$, and the mode cannot decay~\cite{Perczel}. While the linewidth is non-zero for a finite lattice, it is still much smaller than the linewidths of the uniform eigenmodes for $d\lesssim \lambda/\sqrt{2}$. This linewidth depends strongly on the lattice size. It is easy to show numerically that the scaling with the total atom number $N$ satisfies $\upsilon_P/\gamma\approx N^{-3}$, as, e.g., for the most subradiant mode in 1D atomic chains~\cite{Asenjo-Garcia2017a,Zhang2018}. The uniform modes at ${\bf q}=0$ are not protected by momentum conservation; in the limit of an infinite array the uniform in-plane eigenmode only radiates exactly normal to the plane (the zeroth order Bragg peak) for any subwavelength lattice spacing~\cite{CAIT,Facchinetti16,Javanainen19}. For the $\Pc_P$ mode, however, the dipoles oscillate along the $x$ axis and so emit mainly in the lattice plane, where light is absorbed by other atoms and must undergo many scattering events to escape. This leads to this mode also being very subradiant for larger lattices, scaling as $\upsilon_P/\gamma\approx N^{-0.9}$ \cite{Facchinetti16}, with the mode becoming completely dark in the infinite lattice limit, $\lim_{N\rightarrow\infty} \upsilon_P =0$. The in-plane mode $\Pc_I$ has a linewidth which is well described for $d<\lambda$ by its large-$N$ limit ~\cite{CAIT,Facchinetti18} 
\begin{equation}
\label{eq:inplanelinewidth}
\upsilon_{I,\infty}\equiv \lim_{N\rightarrow\infty} \upsilon_I = {3\lambda^2\gamma\over 4\pi d^2}.
\end{equation}

Here we show that these eigenmodes make subradiance-protected excitation spreading possible in 2D arrays. As a consequence, even localized initial excitations can produce collimated emission of light originating from coherently 
oscillating atomic dipoles across the entire lattice. The protocol requires that we choose the initial excitation to match one of these eigenmodes in the center of the lattice, and to be zero everywhere else, such that the excitation has a non-zero overlap with the targeted collective mode. Since this mode is strongly subradiant, and as other modes decay, it will come to dominate and the initial localized excitation will quickly evolve into a coherent, subradiant state extended across the entire lattice. We determine the occupation of a particular eigenmode ${\bf v}_j$ by $L_j(t)$ defined by Eq.~(\ref{eq:measure}).
We take this measure to be normalized at $t=0$.

\subsubsection{Time evolution}

\begin{figure}
\centering
		\includegraphics[width=1\columnwidth]{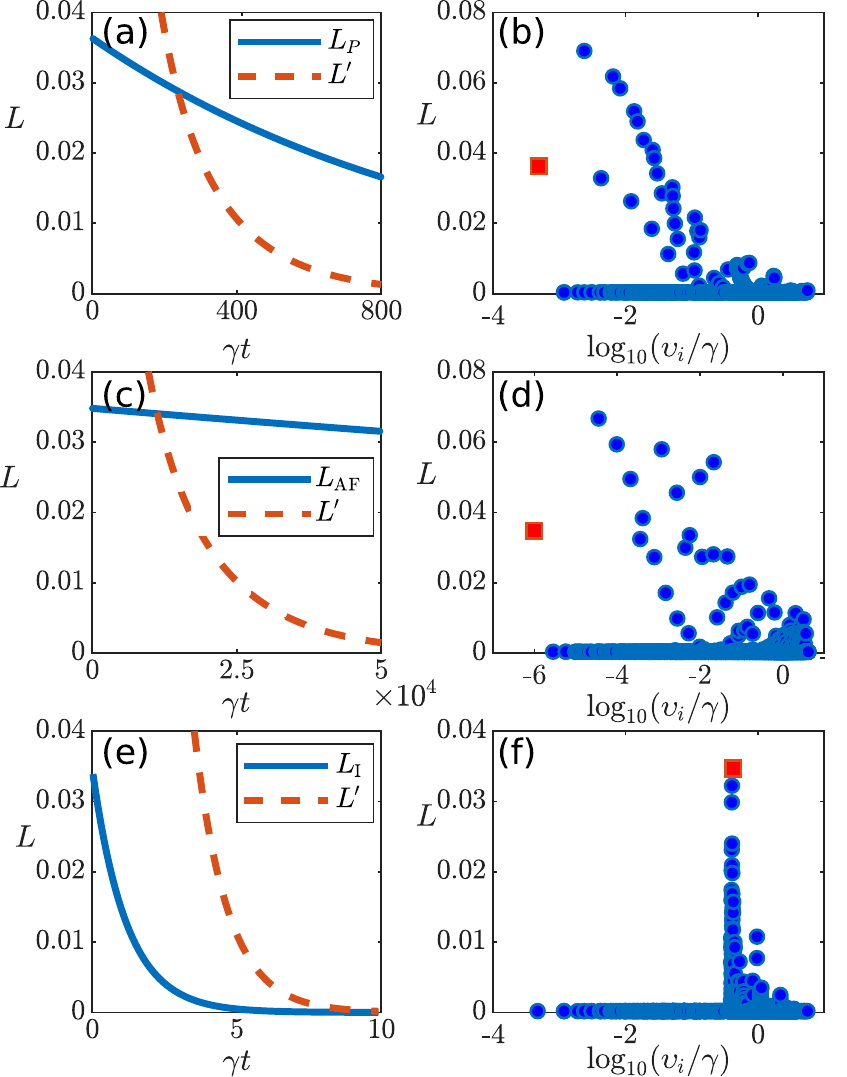}
    \vspace*{-12pt}
		\caption{
		Time evolution and initial eigenmode distribution of a localized excitation. (a) The occupation measure $L_P$ for the uniform out-of-plane mode and the sum $L^\prime$ of the occupation measure of all other modes for an initial in-phase out-of-plane excitation localized on the central nine atoms of a $31\times 31$ lattice, with lattice spacing $d=0.75\lambda$. The delocalized uniform mode quickly grows to $100\%$ of the remaining excitation. (b) The initial mode occupations for the initial excitation and parameters in (a), ordered by collective linewidth $\upsilon_i$. The target uniform out-of-plane mode is illustrated by the red square. (c) Mode occupation $L_\mathrm{AF}$ of the AFM out-of-plane mode and the sum $L^\prime$ of all other mode occupations for an initial out-of-plane excitation localized on the central nine atoms, the phase of which varies by $\pi$ between nearest neighbors. Here the the $31\times 31$ lattice has lattice spacing $d=0.45\lambda$. (d) The initial mode occupations for the initial excitation and parameters in (c), where the red square now denotes the target AFM delocalized mode. (e) Mode occupation $L_I$ of the uniform in-plane mode, and the sum of $L^\prime$ of all other modes for an initial in-phase in-plane excitation on the central nine atoms, showing that in the absence of subradiance the delocalization is not achieved. (f) Initial mode occupation for the excitation in (e), where the red square indicates the uniform in-plane mode. 
		}
		\label{fig:timeandl}
\end{figure}

\begin{figure*}
\centering
		\includegraphics[width=1.8\columnwidth]{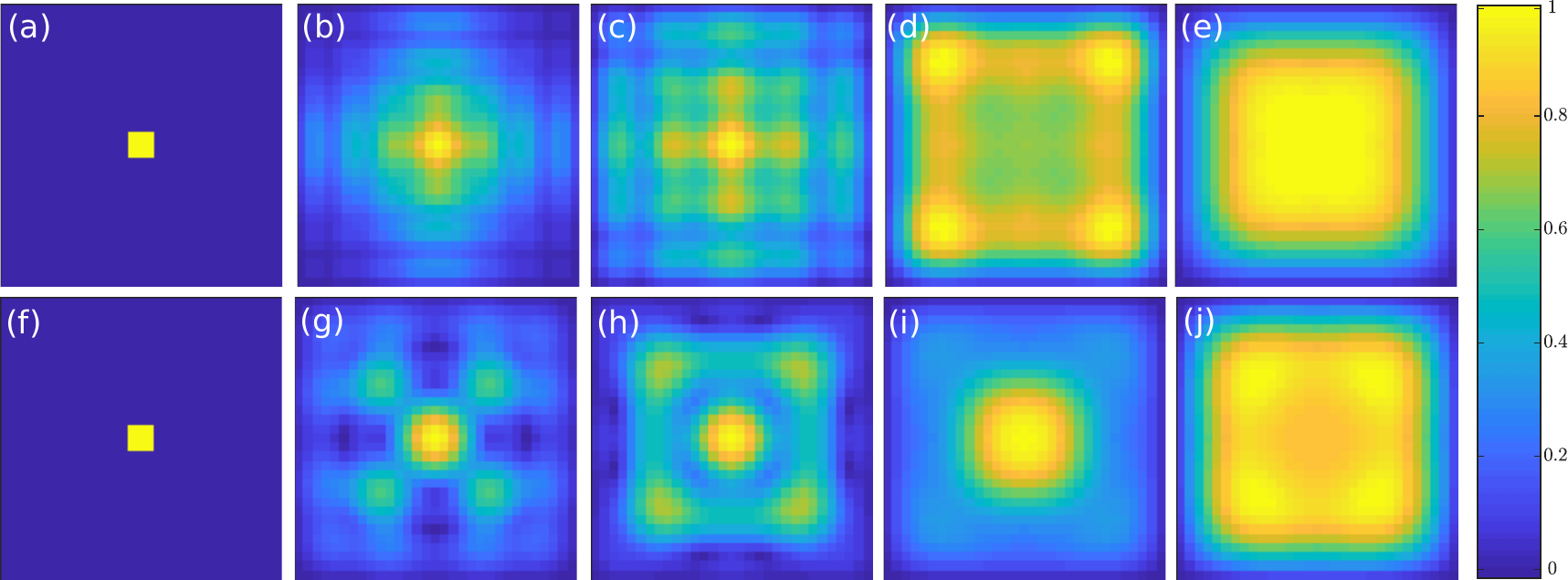}
    \vspace*{-12pt}
		\caption{
		Spatial distribution of the absolute value of polarization $|P|$ (the polarization is oriented normal to the lattice plane) on each atom of a $31\times 31$ lattice at varying times. (a-e) Evolution of an in-phase excitation (targeting uniform $\Pc_P$ mode) with lattice spacing $d=0.75\lambda$ at times $\gamma t=0,50,100,400,$ and $800$, and (f-j) evolution of an AFM excitation (targeting AFM $\Pc_{\mathrm{AF}}$ mode) with lattice spacing $d=0.45\lambda$ at times $\gamma t=0$, $4\times 10^{3}$, $8\times 10^{3}$, $3\times 10^{5}$, and $6\times 10^{5}$, from left to right. Note that at each time slice $|P|$ is normalized independently. 
    }
    \vspace*{-11pt}
		\label{fig:spreading}
\end{figure*}

For a $31\times 31$ square lattice with a lattice spacing of $d=0.75\lambda$ and $\Delta=0$, the uniform mode $\Pc_P$ is very subradiant, with a linewidth $(5\times 10^{-4})\gamma$, as shown in Fig.~\ref{fig:linewidthvspacing}. To target this mode, we start from an initial excitation which has $\mathcal{P}_x^{(j)}=1/3$ on the central nine atoms and $0$ on all other atoms. The resulting time dynamics are shown in Fig.~\ref{fig:timeandl} (a), where the mode occupation of the uniform perpendicular mode, $L_P$, is plotted along with the sum of the occupations of all other eigenmodes,  denoted by $L^\prime$. The dynamics can be understood by looking at the initial mode occupations, plotted in Fig.~\ref{fig:timeandl} (b). While several modes are initially occupied, the target mode, indicated by the red square, is the most subradiant. As the less subradiant modes decay rapidly, the total excitation initially falls off quickly. However, because the target mode is very subradiant, the population of this mode decays much slower, and after some time it becomes the dominant mode. At this point, light is effectively stored in a very subradiant state delocalized across the entire lattice. 

The initial excitation and the time dynamics for the AFM mode are shown in Fig.~\ref{fig:timeandl} (c) and (d). As in the case of the uniform mode, the initial localized excitation tends to a delocalized state. However, due to the sharp transition to a highly subradiant mode at $d<\lambda/\sqrt{2}$, there is a larger gap between this mode and other occupied modes, as can be seen in Fig.~\ref{fig:timeandl} (d). This means that the target delocalized mode decays much less in the time it takes to become the dominant mode. 

The importance of subradiance in letting the excitation spread across the lattice is illustrated by comparing an attempt to target the uniform in-plane mode $\Pc_I$. The result is shown in Fig.~\ref{fig:timeandl} (e) and (f). Because the excited modes are not significantly subradiant, the occupations decay on a much faster timescale. Even within this short time, however, the occupation of the target mode is never higher than that of all other modes.

The spatial spreading is illustrated in Fig.~\ref{fig:spreading}, which shows the absolute value of the excitation amplitude (which always points normal to the plane) across the lattice at various times, for initial in-phase and AFM excitation at the center. For the out-of-plane mode the excitation quickly spreads across the lattice as each eigenmode component of the initial excitation decays at different rates. After about $t=500/\gamma$, the excitation has settled into the most subradiant target mode. While the AFM mode behaves similarly, the timescale is longer due to the smaller linewidth of many of the occupied modes.  

\begin{figure}
\centering
		\includegraphics[width=1\columnwidth]{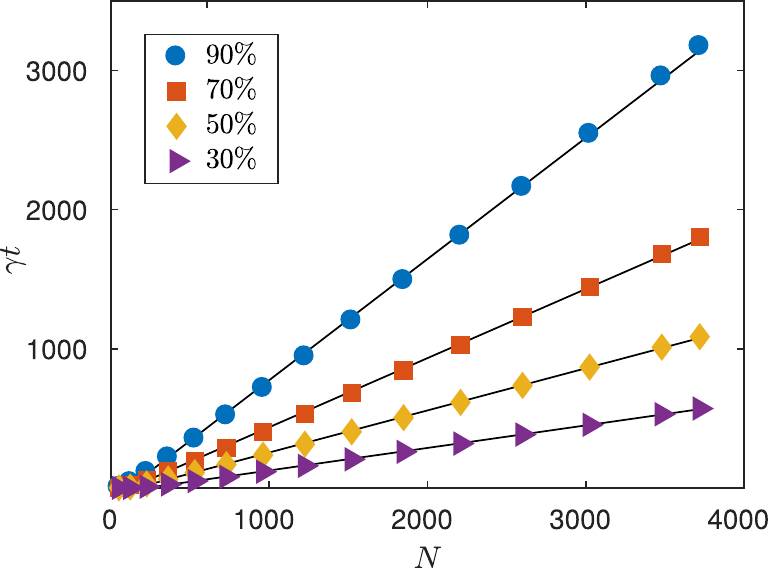}
    \vspace*{-12pt}
		\caption{
		Time $t$ taken for the uniform delocalized mode occupation measure to contribute different percentages of the sum of all occupation measures, as a function of atom number $N$. Black lines show linear fits.  
		}
		\label{fig:tvn}
\end{figure}

In Fig.~\ref{fig:tvn} we plot the time taken for the occupation measure of the target uniform mode to reach a fixed percentage of the sum of all occupations, starting from an initial localized in-phase excitation on the central nine atoms. For each percentage, the time taken is directly proportional to the total atom number $N\sim 1/\gamma_P$ (while for the AFM mode it is proportional to $N^3$). Conversely, we note that the share of the initial occupation measure at $t=0$ is proportional to $1/N$. 

In the case of both the modes $\Pc_P$ and $\Pc_\mathrm{AF}$, while the initial excitation is successfully transferred to the target mode, the final overall occupation is much less than the total initial occupation, which is normalized to one. This is due to the low occupation of the target mode in the initial state, which is approximately $0.035$ in both cases. While the uniform mode $\Pc_p$ is subradiant, the initial occupation still decays to $L\approx 0.018$ by the time the occupation of all other modes has become negligible. The sharp decrease in the lifetime of the AFM mode below the light line leads to a much smaller decay, with $L\approx 0.032$ by the time it has become the dominant mode. 
In the case of a single photon excitation, the occupation measure indicates the probability that the photon has reached the target mode (in the latter case, approximately once in every 30 realizations).
The relatively small initial occupation of both modes is a direct result of the initial excitation being localized to $\approx 1\%$ of the atoms. This could be improved by starting with a larger excitation. The occupation of the target mode rises to $0.07$ for an initial excitation of the central 16 atoms for example, and to $0.1$ for an initial excitation of the central 25 atoms.

\subsection{Photon release}
\label{sub:release}

\begin{figure}
\centering
		\includegraphics[width=1\columnwidth]{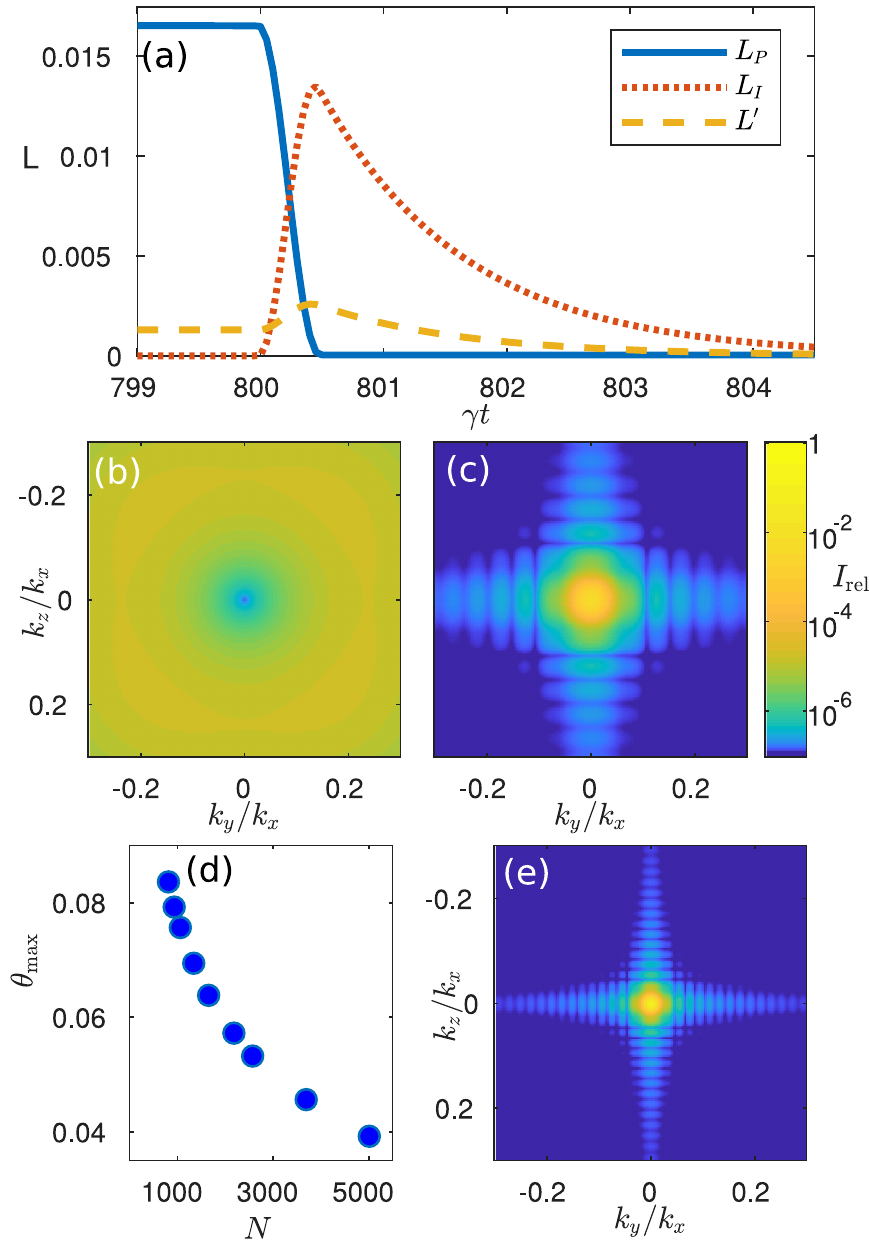}
    \vspace*{-12pt}
		\caption{
		Highly collimated light emission from a uniform subradiant eigenmode delocalized across the entire $31\times 31$ array, with $d=0.75\lambda$. (a) To release the light, Zeeman level splitting $\Delta=3.3\gamma$ is introduced, either by a magnetic field or by AC-stark shifts, at $t=800/\gamma$  for a short period of time $0.5/\gamma$ to transfer the excitation from the collective eigenmode where the dipoles are uniformly pointing normal to the lattice (with the occupation measure $L_P$) to the one where they are coherently oscillating in phase on the lattice plane ($L_I$). The sum of the occupations of all the other modes is $L^\prime$. (b) Image of the far-field radiation of the initial excitation at $t=0$, and (c) the released radiation at $t=801/\gamma$. (d) The angle $\theta_\mathrm{max}$ such that $99\%$ of the integrated far field intensity is between $0<\theta<\theta_\mathrm{max}$, where $\theta=\arctan{(\sqrt{k_y^2+k_z^2}/k_x)}$, as a function of the number of atoms $N$. To compare different lattice sizes, the excitation is released when the target mode accounts for $90\%$ of the population. (e) Far-field radiation for a $71 \times 71$ lattice, released when $90\%$ of the excitation was in the mode $\Pc_P$, showing highly directional emission of the stored excitation. The intensity $I_\mathrm{rel}$ is plotted in relative units scaled to a maximum of one. }
		\label{fig:detuning}
\end{figure}

In the previous section we described how an initially localized excitation can be transferred to a delocalized subradiant eigenmode occupation, extending over the entire array of atoms.
As the photon then is effectively stored in a subradiant state, it decays slowly and little light is emitted. To release the photon, the Zeeman splitting can be introduced at the desired time by turning on a magnetic field or by utilizing AC-Stark shifts of lasers or microwaves~\cite{gerbier_pra_2006}. As discussed in Sec.~\ref{ss:single_atom} for a single-atom, this couples the $x$ and $y$ components of polarization.

The effect on the many-atom lattice can similarly be understood by considering the uniform in-plane eigenmode $\Pc_I$, and one of the two out-of-plane eigenmodes of the system $\Pc_\mathrm{AF}$ and $\Pc_p$. When we introduce a non-zero Zeeman splitting, these modes are no longer eigenmodes, but coupling is introduced between in-plane and out-of-plane modes. The dynamics in the presence of Zeeman splitting can be understood by considering a two-mode model~\cite{Facchinetti16,Facchinetti18}, which for a driven system represents a linearized version of the EIT equations for bright and dark states~\cite{FleischhauerEtAlRMP2005}. The two-mode model qualitatively captures many aspects of the dynamics of the full lattice for sufficiently large arrays, since the phase-matching conditions of the other modes are not satisfied. 
We find that it can also illustrate how the excitation is transferred from one of the dark, subradiant modes to the uniform in-plane mode in the present case. 
In analogy to the case for a single atom described by Eq.~(\ref{zeeman1}),~(\ref{zeeman2}), the relevant physics is captured by the simple coupled two-mode dynamics 
\begin{subequations}
\begin{align}
\dot\Pc_{P\backslash\mathrm{AF}} & = (i \delta_{P\backslash\mathrm{AF}} -\upsilon_{P\backslash\mathrm{AF}}) \Pc_{P\backslash\mathrm{AF}} - \Delta^{(j)}  \Pc_I, \label{coll2modex}\\
\dot \Pc_I & = (i \delta_I -\upsilon_I) \Pc_I +  \Delta^{(j)}  \Pc_{P\backslash\mathrm{AF}} \, ,
\label{coll2modey}
\end{align} \label{bothtwo}
\end{subequations} 
where $\delta_{P\backslash\mathrm{AF}}$ is the collective resonance shift and $\upsilon_{P\backslash\mathrm{AF}}$ the collective linewidth of the uniform out-of-plane or AFM out-of-plane mode, and $\delta_I$ the collective resonance shift and $\upsilon_I$ the collective linewidth of the uniform in-plane mode. Here to couple $\Pc_p$ and $\Pc_I$ we choose the symmetric Zeeman splitting on atom $j$ to be $\Delta^{(j)}=\Delta$, identical for all atoms, while to couple $\Pc_\mathrm{AF}$ and $\Pc_I$ we take the splitting on atom $j$, indexed in the $y$ and $z$ direction by $j_y, j_z$ respectively, to be $\Delta^{(j)}=(-1)^{j_y}(-1)^{j_z}\Delta$, alternating between neighboring sites. 

As seen from Eq.~(\ref{coll2modex}) and~(\ref{coll2modey}), in the presence of Zeeman splitting, $\Pc_P$ or $\Pc_\mathrm{AF}$ is no longer an eigenmode, and the dipoles start to rotate towards the plane. Applying the splitting for a short time, the excitation can then be transferred to the mode $\Pc_I$, where each atom has approximately uniform in-phase polarization in the $y$ direction, with much faster emission rate, allowing the excitation to quickly radiate away. We calculate the emitted light, given by the sum of all the atomic contributions from Eq.~(\ref{eq:e_field}),  in the far-field limit ($r\gg d,\lambda$, where $r$ is the distance from the source to the observation point)~\cite{Jackson}. While we show here numerical results for the case of the uniform mode, the stored AFM mode could also be released in a similar manner. The two mode model captures much of the physics because once the dipoles are all oscillating with a fixed phase relation, they will continue to do so even when Zeeman splitting is present and the direction of the effective magnetic field breaks the isotropy of the $J=0\rightarrow J'=1$ transition.


Figure~\ref{fig:detuning} shows the results of Zeeman splitting being turned on at $t=800/\gamma$ for a short time, for the stored uniform out-of-plane mode. As soon as the splitting is turned on, the occupation $\Pc_P$ falls rapidly, while the occupation of the in-plane mode $\Pc_I$ shows a corresponding rise. Since this mode has a much larger linewidth, the occupation then begins to fall quickly and, within a time $\approx 5/\gamma$, the photon has been emitted from the lattice.
 
While emission from the initial excitation is omnidirectional, emission from the delocalized mode is highly collimated along the direction of the $x$ axis, perpendicular to the lattice. We quantify this by the angle $\theta_\mathrm{max}$ such that integrating between $0\leq\theta\leq \theta_\mathrm{max}$ and $0\leq \phi\leq 2\pi$ gives $99\%$ of the total integrated forward-scattered intensity, where $\theta=\arctan{(\sqrt{k_y^2+k_z^2}/k_x)}$ is the polar angle of the ray to the $k_x$ axis and $\phi=\arctan{(k_y/k_z)}$ the azimuthal angle in the $k_y k_z$ plane. For the initial excitation, this angle is $\theta_\mathrm{max}= 0.997 (\pi/2)$, i.e.~the light is not collimated at all.  However, when the excitation is released from the delocalized mode, the photon emission is highly directional, with $\theta_\mathrm{max}=0.05(\pi/2)$. In this case the emission is equal in the forward and backward directions. Forward-only scattering could be achieved using two arrays offset in the $x$ direction with a suitable phase shift~\cite{Grankin18}, analogously to the directed radiation of antennas. In the presence of resonant-coupled light and atoms, each layer can be considered as a `superatom', exhibiting its own resonance linewidth and line shift, and the analysis of the system is similar to the one of two 1D atoms~\cite{Facchinetti18,Yoo18,Javanainen19}. 

The far-field radiation pattern of the delocalized mode is that of a 2D diffraction grating, dominated by the central zeroth order Bragg peak (the higher order Bragg peaks do not exist because of the subwavelength lattice spacing). For lattices with a higher number of atoms, this central peak becomes sharper. To compare different lattice sizes we start with the same initial excitation on the central nine atoms, and let it evolve until the target uniform mode accounts for $90\%$ of the remaining mode occupation. Increasing the number of atoms leads to a sharp drop in $\theta_\mathrm{max}$, as shown in Fig.~\ref{fig:detuning} (d). For the largest lattice ($71\times 71$), we find $\theta_\mathrm{max}=0.02(\pi/2)$ which represents a photon wave-packet that is highly localized in $k$-space. For an infinite array, the propagation reaches the precise 1D limit, propagating only in the $x$ direction.

The process described in this section is reminiscent of the procedure to slow and store light within an atomic cloud using the standard single-particle EIT~\cite{LiuEtAlNature2001,DuttonHau}. In our case it is the collective modes of the atomic array which act as the bright and dark states. To release the photon, the Zeeman splitting couples these states, playing the role of the coupling laser which is usually used to restore transparency. This allows the photon to radiate away while preserving spatial coherence.

\subsection{The effects of position fluctuations}

\begin{figure}
\centering
		\includegraphics[width=1\columnwidth]{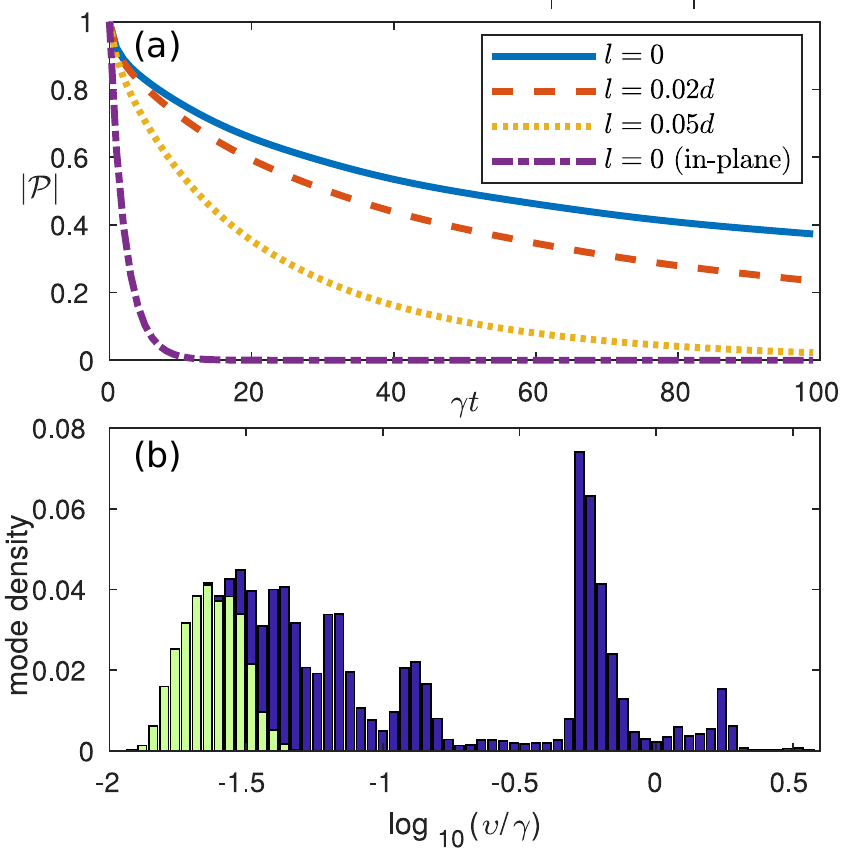}
    \vspace*{-12pt}
		\caption{ 
		(a) Decay of polarization $|P|$ for a localized in-phase uniform out-of-plane excitation for position fluctuations with standard deviation $l$ around each site, and of a localized in-phase in-plane excitation at fixed atom position for comparison. At long times, the out-of-plane excitations are well described by a single exponential, and the extracted decay rates $\gamma_e$ for $\gamma t>40$, from top to bottom, are $\gamma_e/\gamma=6\times10^{-3}$, $0.01$, and $0.03$. The decay rate of the in-plane excitation is $0.5\gamma$. (b) A histogram of the occupation measure of the initial excitation of in-phase polarization on the central nine atoms with each eigenmode having a decay rate $\upsilon$ (dark blue), and the distribution of the most subradiant modes which contribute up to $30\%$ of the excitation for each realizations (light green). Results are averaged over 5000 individual realizations of position fluctuations. 
		}
		\label{fig:disorder}
\end{figure}

While the depth of the atomic lattice potential is sufficient to ensure the atoms remain in the ground state of the trap, the wave-function in this state has a finite size. We account for this by taking many individual realizations with fixed atom positions drawn randomly from a  harmonic oscillator ground-state probability distribution for each lattice site with root-mean-square width $l$, and stochastically averaging over these realizations~\cite{Jenkins2012a}. This procedure has been shown to be exact in the limit of low light intensity~\cite{Javanainen1999a,Lee16}. The result of spatial disorder is shown in Fig.~\ref{fig:disorder}, where we plot the surviving amplitude $\Pc=\sqrt{\sum_{j,\mu}|\Pc_{\mu}^{(j)}|^2}$ as a function of time for varying fluctuation lengths. For increasing disorder, the excitation decays more quickly, as also observed in other subradiance-protected excitation transfer studies~\cite{Needham19}. However, in these cases the lifetime is still much longer than the corresponding case of an in-phase in-plane localized excitation with no disorder. This can be understood by looking at the  distribution of the eigenmodes, weighted by their initial occupation $L$, across many stochastic realizations. This is compared in Fig.~\ref{fig:disorder} in the case of the uniform excitation to the distribution of the most subradiant modes which contribute up to $30\%$ of the initial excitation of each realization, shown in green. Although disorder means that each realization has far fewer very subradiant modes, these few modes are consistently well-represented in the initial excitation.		

\section{Concluding remarks}
\label{sec:conc}

Subradiant states are isolated from the environment and therefore difficult to excite. Standard field excitation typically only results in a very small fraction of the total population to notably subradiant modes~\cite{Guerin_subr16}. Transferring a more substantial population to very slowly radiating states typically requires first the breaking of the eigenmode symmetry before the excitation, followed by restoration of the symmetry, such that the modes are temporarily made to interact with radiation~\cite{Facchinetti16}. 

Here we showed how to achieve subradiance-protected excitation spreading in 2D arrays by manipulating the atomic level shifts. Remarkably, this allows us to use even an initial localized single-photon excitation to produce a delocalized collimated emission from the array. 
In the procedure a localized single-photon excitation of an atomic lattice, with omnidirectional emission, spreads out into a subradiant state which is delocalized across the whole lattice and is protected from decay. 

We considered two strongly subradiant target eigenmodes into which the initial excitation can evolve and showed how they can be excited. In the both cases the suppressed emission is achieved by reorienting the atomic dipoles normal to the lattice plane. 
For the uniformly oscillating dipoles, the mode is similar to the one studied in Refs.~\cite{Facchinetti16,Facchinetti18} for coherent driving. The AFM mode, however, with the polarization of each atom $\pi$ out of phase with its nearest neighbors, can be significantly more subradiant, as it crosses the light line at $d=\lambda/\sqrt{2}$~\cite{Perczel}, and is 
particularly promising for potential applications involving a photon storage.

Our protocol could form part of a quantum information or quantum computing architecture~\cite{Grankin18,Duan10,Northup2014}, coupling via short-range interactions to an input state, and coherently converting this local excitation into directional emission, and effective 1D propagation. This output could then be transferred via free space to another stage. Future work could identify collective quantum effects in higher-order correlations beyond the single-excitation limit, and study the effect of such correlations on the emitted collimated light.

\begin{acknowledgments}
We thank Chris Parmee for reading and commenting on the manuscript.
We acknowledge financial support from  EPSRC.
\end{acknowledgments}


\begin{thebibliography}{85}%
\makeatletter
\providecommand \@ifxundefined [1]{%
 \@ifx{#1\undefined}
}%
\providecommand \@ifnum [1]{%
 \ifnum #1\expandafter \@firstoftwo
 \else \expandafter \@secondoftwo
 \fi
}%
\providecommand \@ifx [1]{%
 \ifx #1\expandafter \@firstoftwo
 \else \expandafter \@secondoftwo
 \fi
}%
\providecommand \natexlab [1]{#1}%
\providecommand \enquote  [1]{``#1''}%
\providecommand \bibnamefont  [1]{#1}%
\providecommand \bibfnamefont [1]{#1}%
\providecommand \citenamefont [1]{#1}%
\providecommand \href@noop [0]{\@secondoftwo}%
\providecommand \href [0]{\begingroup \@sanitize@url \@href}%
\providecommand \@href[1]{\@@startlink{#1}\@@href}%
\providecommand \@@href[1]{\endgroup#1\@@endlink}%
\providecommand \@sanitize@url [0]{\catcode `\\12\catcode `\$12\catcode
  `\&12\catcode `\#12\catcode `\^12\catcode `\_12\catcode `\%12\relax}%
\providecommand \@@startlink[1]{}%
\providecommand \@@endlink[0]{}%
\providecommand \url  [0]{\begingroup\@sanitize@url \@url }%
\providecommand \@url [1]{\endgroup\@href {#1}{\urlprefix }}%
\providecommand \urlprefix  [0]{URL }%
\providecommand \Eprint [0]{\href }%
\providecommand \doibase [0]{http://dx.doi.org/}%
\providecommand \selectlanguage [0]{\@gobble}%
\providecommand \bibinfo  [0]{\@secondoftwo}%
\providecommand \bibfield  [0]{\@secondoftwo}%
\providecommand \translation [1]{[#1]}%
\providecommand \BibitemOpen [0]{}%
\providecommand \bibitemStop [0]{}%
\providecommand \bibitemNoStop [0]{.\EOS\space}%
\providecommand \EOS [0]{\spacefactor3000\relax}%
\providecommand \BibitemShut  [1]{\csname bibitem#1\endcsname}%
\let\auto@bib@innerbib\@empty
\bibitem [{\citenamefont {Balik}\ \emph {et~al.}(2013)\citenamefont {Balik},
  \citenamefont {Win}, \citenamefont {Havey}, \citenamefont {Sokolov},\ and\
  \citenamefont {Kupriyanov}}]{BalikEtAl2013}%
  \BibitemOpen
  \bibfield  {author} {\bibinfo {author} {\bibfnamefont {S.}~\bibnamefont
  {Balik}}, \bibinfo {author} {\bibfnamefont {A.~L.}\ \bibnamefont {Win}},
  \bibinfo {author} {\bibfnamefont {M.~D.}\ \bibnamefont {Havey}}, \bibinfo
  {author} {\bibfnamefont {I.~M.}\ \bibnamefont {Sokolov}}, \ and\ \bibinfo
  {author} {\bibfnamefont {D.~V.}\ \bibnamefont {Kupriyanov}},\ }\bibfield
  {title} {\enquote {\bibinfo {title} {Near-resonance light scattering from a
  high-density ultracold atomic ${}^{87}${Rb} gas},}\ }\href {\doibase
  10.1103/PhysRevA.87.053817} {\bibfield  {journal} {\bibinfo  {journal} {Phys.
  Rev. A}\ }\textbf {\bibinfo {volume} {87}},\ \bibinfo {pages} {053817}
  (\bibinfo {year} {2013})}\BibitemShut {NoStop}%
\bibitem [{\citenamefont {Chab\'e}\ \emph {et~al.}(2014)\citenamefont
  {Chab\'e}, \citenamefont {Rouabah}, \citenamefont {Bellando}, \citenamefont
  {Bienaim\'e}, \citenamefont {Piovella}, \citenamefont {Bachelard},\ and\
  \citenamefont {Kaiser}}]{CHA14}%
  \BibitemOpen
  \bibfield  {author} {\bibinfo {author} {\bibfnamefont {Julien}\ \bibnamefont
  {Chab\'e}}, \bibinfo {author} {\bibfnamefont {Mohamed-Taha}\ \bibnamefont
  {Rouabah}}, \bibinfo {author} {\bibfnamefont {Louis}\ \bibnamefont
  {Bellando}}, \bibinfo {author} {\bibfnamefont {Tom}\ \bibnamefont
  {Bienaim\'e}}, \bibinfo {author} {\bibfnamefont {Nicola}\ \bibnamefont
  {Piovella}}, \bibinfo {author} {\bibfnamefont {Romain}\ \bibnamefont
  {Bachelard}}, \ and\ \bibinfo {author} {\bibfnamefont {Robin}\ \bibnamefont
  {Kaiser}},\ }\bibfield  {title} {\enquote {\bibinfo {title} {Coherent and
  incoherent multiple scattering},}\ }\href {\doibase
  10.1103/PhysRevA.89.043833} {\bibfield  {journal} {\bibinfo  {journal} {Phys.
  Rev. A}\ }\textbf {\bibinfo {volume} {89}},\ \bibinfo {pages} {043833}
  (\bibinfo {year} {2014})}\BibitemShut {NoStop}%
\bibitem [{\citenamefont {Pellegrino}\ \emph {et~al.}(2014)\citenamefont
  {Pellegrino}, \citenamefont {Bourgain}, \citenamefont {Jennewein},
  \citenamefont {Sortais}, \citenamefont {Browaeys}, \citenamefont {Jenkins},\
  and\ \citenamefont {Ruostekoski}}]{Pellegrino2014a}%
  \BibitemOpen
  \bibfield  {author} {\bibinfo {author} {\bibfnamefont {J.}~\bibnamefont
  {Pellegrino}}, \bibinfo {author} {\bibfnamefont {R.}~\bibnamefont
  {Bourgain}}, \bibinfo {author} {\bibfnamefont {S.}~\bibnamefont {Jennewein}},
  \bibinfo {author} {\bibfnamefont {Y.~R.~P.}\ \bibnamefont {Sortais}},
  \bibinfo {author} {\bibfnamefont {A.}~\bibnamefont {Browaeys}}, \bibinfo
  {author} {\bibfnamefont {S.~D.}\ \bibnamefont {Jenkins}}, \ and\ \bibinfo
  {author} {\bibfnamefont {J.}~\bibnamefont {Ruostekoski}},\ }\bibfield
  {title} {\enquote {\bibinfo {title} {Observation of suppression of light
  scattering induced by dipole-dipole interactions in a cold-atom ensemble},}\
  }\href {\doibase 10.1103/PhysRevLett.113.133602} {\bibfield  {journal}
  {\bibinfo  {journal} {Phys. Rev. Lett.}\ }\textbf {\bibinfo {volume} {113}},\
  \bibinfo {pages} {133602} (\bibinfo {year} {2014})}\BibitemShut {NoStop}%
\bibitem [{\citenamefont {Sheremet}\ \emph {et~al.}(2014)\citenamefont
  {Sheremet}, \citenamefont {Sokolov}, \citenamefont {Kupriyanov},
  \citenamefont {Balik}, \citenamefont {Win},\ and\ \citenamefont
  {Havey}}]{Havey_jmo14}%
  \BibitemOpen
  \bibfield  {author} {\bibinfo {author} {\bibfnamefont {A.S.}\ \bibnamefont
  {Sheremet}}, \bibinfo {author} {\bibfnamefont {I.M.}\ \bibnamefont
  {Sokolov}}, \bibinfo {author} {\bibfnamefont {D.V.}\ \bibnamefont
  {Kupriyanov}}, \bibinfo {author} {\bibfnamefont {S.}~\bibnamefont {Balik}},
  \bibinfo {author} {\bibfnamefont {A.L.}\ \bibnamefont {Win}}, \ and\ \bibinfo
  {author} {\bibfnamefont {M.D.}\ \bibnamefont {Havey}},\ }\bibfield  {title}
  {\enquote {\bibinfo {title} {Light scattering on the {$F = 1 \rightarrow F' =
  0$} transition in a cold and high density {$^{87}$Rb} vapor},}\ }\href
  {\doibase 10.1080/09500340.2013.854419} {\bibfield  {journal} {\bibinfo
  {journal} {Journal of Modern Optics}\ }\textbf {\bibinfo {volume} {61}},\
  \bibinfo {pages} {77--84} (\bibinfo {year} {2014})}\BibitemShut {NoStop}%
\bibitem [{\citenamefont {Kwong}\ \emph {et~al.}(2014)\citenamefont {Kwong},
  \citenamefont {Yang}, \citenamefont {Pramod}, \citenamefont {Pandey},
  \citenamefont {Delande}, \citenamefont {Pierrat},\ and\ \citenamefont
  {Wilkowski}}]{wilkowski}%
  \BibitemOpen
  \bibfield  {author} {\bibinfo {author} {\bibfnamefont {C.~C.}\ \bibnamefont
  {Kwong}}, \bibinfo {author} {\bibfnamefont {T.}~\bibnamefont {Yang}},
  \bibinfo {author} {\bibfnamefont {M.~S.}\ \bibnamefont {Pramod}}, \bibinfo
  {author} {\bibfnamefont {K.}~\bibnamefont {Pandey}}, \bibinfo {author}
  {\bibfnamefont {D.}~\bibnamefont {Delande}}, \bibinfo {author} {\bibfnamefont
  {R.}~\bibnamefont {Pierrat}}, \ and\ \bibinfo {author} {\bibfnamefont
  {D.}~\bibnamefont {Wilkowski}},\ }\bibfield  {title} {\enquote {\bibinfo
  {title} {Cooperative emission of a coherent superflash of light},}\ }\href
  {\doibase 10.1103/PhysRevLett.113.223601} {\bibfield  {journal} {\bibinfo
  {journal} {Phys. Rev. Lett.}\ }\textbf {\bibinfo {volume} {113}},\ \bibinfo
  {pages} {223601} (\bibinfo {year} {2014})}\BibitemShut {NoStop}%
\bibitem [{\citenamefont {Jennewein}\ \emph {et~al.}(2016)\citenamefont
  {Jennewein}, \citenamefont {Besbes}, \citenamefont {Schilder}, \citenamefont
  {Jenkins}, \citenamefont {Sauvan}, \citenamefont {Ruostekoski}, \citenamefont
  {Greffet}, \citenamefont {Sortais},\ and\ \citenamefont
  {Browaeys}}]{Jennewein_trans}%
  \BibitemOpen
  \bibfield  {author} {\bibinfo {author} {\bibfnamefont {S.}~\bibnamefont
  {Jennewein}}, \bibinfo {author} {\bibfnamefont {M.}~\bibnamefont {Besbes}},
  \bibinfo {author} {\bibfnamefont {N.~J.}\ \bibnamefont {Schilder}}, \bibinfo
  {author} {\bibfnamefont {S.~D.}\ \bibnamefont {Jenkins}}, \bibinfo {author}
  {\bibfnamefont {C.}~\bibnamefont {Sauvan}}, \bibinfo {author} {\bibfnamefont
  {J.}~\bibnamefont {Ruostekoski}}, \bibinfo {author} {\bibfnamefont {J.-J.}\
  \bibnamefont {Greffet}}, \bibinfo {author} {\bibfnamefont {Y.~R.~P.}\
  \bibnamefont {Sortais}}, \ and\ \bibinfo {author} {\bibfnamefont
  {A.}~\bibnamefont {Browaeys}},\ }\bibfield  {title} {\enquote {\bibinfo
  {title} {Coherent scattering of near-resonant light by a dense microscopic
  cold atomic cloud},}\ }\href {\doibase 10.1103/PhysRevLett.116.233601}
  {\bibfield  {journal} {\bibinfo  {journal} {Phys. Rev. Lett.}\ }\textbf
  {\bibinfo {volume} {116}},\ \bibinfo {pages} {233601} (\bibinfo {year}
  {2016})}\BibitemShut {NoStop}%
\bibitem [{\citenamefont {Kwong}\ \emph {et~al.}(2015)\citenamefont {Kwong},
  \citenamefont {Yang}, \citenamefont {Delande}, \citenamefont {Pierrat},\ and\
  \citenamefont {Wilkowski}}]{wilkowski2}%
  \BibitemOpen
  \bibfield  {author} {\bibinfo {author} {\bibfnamefont {C.~C.}\ \bibnamefont
  {Kwong}}, \bibinfo {author} {\bibfnamefont {T.}~\bibnamefont {Yang}},
  \bibinfo {author} {\bibfnamefont {D.}~\bibnamefont {Delande}}, \bibinfo
  {author} {\bibfnamefont {R.}~\bibnamefont {Pierrat}}, \ and\ \bibinfo
  {author} {\bibfnamefont {D.}~\bibnamefont {Wilkowski}},\ }\bibfield  {title}
  {\enquote {\bibinfo {title} {Cooperative emission of a pulse train in an
  optically thick scattering medium},}\ }\href {\doibase
  10.1103/PhysRevLett.115.223601} {\bibfield  {journal} {\bibinfo  {journal}
  {Phys. Rev. Lett.}\ }\textbf {\bibinfo {volume} {115}},\ \bibinfo {pages}
  {223601} (\bibinfo {year} {2015})}\BibitemShut {NoStop}%
\bibitem [{\citenamefont {Bromley}\ \emph {et~al.}(2016)\citenamefont
  {Bromley}, \citenamefont {Zhu}, \citenamefont {Bishof}, \citenamefont
  {Zhang}, \citenamefont {Bothwell}, \citenamefont {Schachenmayer},
  \citenamefont {Nicholson}, \citenamefont {Kaiser}, \citenamefont {Yelin},
  \citenamefont {Lukin}, \citenamefont {Rey},\ and\ \citenamefont
  {Ye}}]{Ye2016}%
  \BibitemOpen
  \bibfield  {author} {\bibinfo {author} {\bibfnamefont {S.~L.}\ \bibnamefont
  {Bromley}}, \bibinfo {author} {\bibfnamefont {B.}~\bibnamefont {Zhu}},
  \bibinfo {author} {\bibfnamefont {M.}~\bibnamefont {Bishof}}, \bibinfo
  {author} {\bibfnamefont {X.}~\bibnamefont {Zhang}}, \bibinfo {author}
  {\bibfnamefont {T.}~\bibnamefont {Bothwell}}, \bibinfo {author}
  {\bibfnamefont {J.}~\bibnamefont {Schachenmayer}}, \bibinfo {author}
  {\bibfnamefont {T.~L.}\ \bibnamefont {Nicholson}}, \bibinfo {author}
  {\bibfnamefont {R.}~\bibnamefont {Kaiser}}, \bibinfo {author} {\bibfnamefont
  {S.~F.}\ \bibnamefont {Yelin}}, \bibinfo {author} {\bibfnamefont {M.~D.}\
  \bibnamefont {Lukin}}, \bibinfo {author} {\bibfnamefont {A.~M.}\ \bibnamefont
  {Rey}}, \ and\ \bibinfo {author} {\bibfnamefont {J.}~\bibnamefont {Ye}},\
  }\bibfield  {title} {\enquote {\bibinfo {title} {Collective atomic scattering
  and motional effects in a dense coherent medium},}\ }\href
  {http://dx.doi.org/10.1038/ncomms11039} {\bibfield  {journal} {\bibinfo
  {journal} {Nat Commun}\ }\textbf {\bibinfo {volume} {7}},\ \bibinfo {pages}
  {11039} (\bibinfo {year} {2016})}\BibitemShut {NoStop}%
\bibitem [{\citenamefont {Jenkins}\ \emph
  {et~al.}(2016{\natexlab{a}})\citenamefont {Jenkins}, \citenamefont
  {Ruostekoski}, \citenamefont {Javanainen}, \citenamefont {Bourgain},
  \citenamefont {Jennewein}, \citenamefont {Sortais},\ and\ \citenamefont
  {Browaeys}}]{Jenkins_thermshift}%
  \BibitemOpen
  \bibfield  {author} {\bibinfo {author} {\bibfnamefont {S.~D.}\ \bibnamefont
  {Jenkins}}, \bibinfo {author} {\bibfnamefont {J.}~\bibnamefont
  {Ruostekoski}}, \bibinfo {author} {\bibfnamefont {J.}~\bibnamefont
  {Javanainen}}, \bibinfo {author} {\bibfnamefont {R.}~\bibnamefont
  {Bourgain}}, \bibinfo {author} {\bibfnamefont {S.}~\bibnamefont {Jennewein}},
  \bibinfo {author} {\bibfnamefont {Y.~R.~P.}\ \bibnamefont {Sortais}}, \ and\
  \bibinfo {author} {\bibfnamefont {A.}~\bibnamefont {Browaeys}},\ }\bibfield
  {title} {\enquote {\bibinfo {title} {Optical resonance shifts in the
  fluorescence of thermal and cold atomic gases},}\ }\href {\doibase
  10.1103/PhysRevLett.116.183601} {\bibfield  {journal} {\bibinfo  {journal}
  {Phys. Rev. Lett.}\ }\textbf {\bibinfo {volume} {116}},\ \bibinfo {pages}
  {183601} (\bibinfo {year} {2016}{\natexlab{a}})}\BibitemShut {NoStop}%
\bibitem [{\citenamefont {Bons}\ \emph {et~al.}(2016)\citenamefont {Bons},
  \citenamefont {de~Haas}, \citenamefont {de~Jong}, \citenamefont {Groot},\
  and\ \citenamefont {van~der Straten}}]{vdStraten16}%
  \BibitemOpen
  \bibfield  {author} {\bibinfo {author} {\bibfnamefont {P.~C.}\ \bibnamefont
  {Bons}}, \bibinfo {author} {\bibfnamefont {R.}~\bibnamefont {de~Haas}},
  \bibinfo {author} {\bibfnamefont {D.}~\bibnamefont {de~Jong}}, \bibinfo
  {author} {\bibfnamefont {A.}~\bibnamefont {Groot}}, \ and\ \bibinfo {author}
  {\bibfnamefont {P.}~\bibnamefont {van~der Straten}},\ }\bibfield  {title}
  {\enquote {\bibinfo {title} {Quantum enhancement of the index of refraction
  in a {Bose-Einstein} condensate},}\ }\href {\doibase
  10.1103/PhysRevLett.116.173602} {\bibfield  {journal} {\bibinfo  {journal}
  {Phys. Rev. Lett.}\ }\textbf {\bibinfo {volume} {116}},\ \bibinfo {pages}
  {173602} (\bibinfo {year} {2016})}\BibitemShut {NoStop}%
\bibitem [{\citenamefont {Guerin}\ \emph {et~al.}(2016)\citenamefont {Guerin},
  \citenamefont {Ara\'ujo},\ and\ \citenamefont {Kaiser}}]{Guerin_subr16}%
  \BibitemOpen
  \bibfield  {author} {\bibinfo {author} {\bibfnamefont {William}\ \bibnamefont
  {Guerin}}, \bibinfo {author} {\bibfnamefont {Michelle~O.}\ \bibnamefont
  {Ara\'ujo}}, \ and\ \bibinfo {author} {\bibfnamefont {Robin}\ \bibnamefont
  {Kaiser}},\ }\bibfield  {title} {\enquote {\bibinfo {title} {Subradiance in a
  large cloud of cold atoms},}\ }\href {\doibase
  10.1103/PhysRevLett.116.083601} {\bibfield  {journal} {\bibinfo  {journal}
  {Phys. Rev. Lett.}\ }\textbf {\bibinfo {volume} {116}},\ \bibinfo {pages}
  {083601} (\bibinfo {year} {2016})}\BibitemShut {NoStop}%
\bibitem [{\citenamefont {Machluf}\ \emph {et~al.}(2019)\citenamefont
  {Machluf}, \citenamefont {Naber}, \citenamefont {Soudijn}, \citenamefont
  {Ruostekoski},\ and\ \citenamefont {Spreeuw}}]{Machluf2018}%
  \BibitemOpen
  \bibfield  {author} {\bibinfo {author} {\bibfnamefont {Shimon}\ \bibnamefont
  {Machluf}}, \bibinfo {author} {\bibfnamefont {Julian~B.}\ \bibnamefont
  {Naber}}, \bibinfo {author} {\bibfnamefont {Maarten~L.}\ \bibnamefont
  {Soudijn}}, \bibinfo {author} {\bibfnamefont {Janne}\ \bibnamefont
  {Ruostekoski}}, \ and\ \bibinfo {author} {\bibfnamefont {Robert J.~C.}\
  \bibnamefont {Spreeuw}},\ }\bibfield  {title} {\enquote {\bibinfo {title}
  {Collective suppression of optical hyperfine pumping in dense clouds of atoms
  in microtraps},}\ }\href {\doibase 10.1103/PhysRevA.100.051801} {\bibfield
  {journal} {\bibinfo  {journal} {Phys. Rev. A}\ }\textbf {\bibinfo {volume}
  {100}},\ \bibinfo {pages} {051801} (\bibinfo {year} {2019})}\BibitemShut
  {NoStop}%
\bibitem [{\citenamefont {Corman}\ \emph {et~al.}(2017)\citenamefont {Corman},
  \citenamefont {Ville}, \citenamefont {Saint-Jalm}, \citenamefont
  {Aidelsburger}, \citenamefont {Bienaim\'e}, \citenamefont {Nascimb\`ene},
  \citenamefont {Dalibard},\ and\ \citenamefont {Beugnon}}]{Dalibard_slab}%
  \BibitemOpen
  \bibfield  {author} {\bibinfo {author} {\bibfnamefont {L.}~\bibnamefont
  {Corman}}, \bibinfo {author} {\bibfnamefont {J.~L.}\ \bibnamefont {Ville}},
  \bibinfo {author} {\bibfnamefont {R.}~\bibnamefont {Saint-Jalm}}, \bibinfo
  {author} {\bibfnamefont {M.}~\bibnamefont {Aidelsburger}}, \bibinfo {author}
  {\bibfnamefont {T.}~\bibnamefont {Bienaim\'e}}, \bibinfo {author}
  {\bibfnamefont {S.}~\bibnamefont {Nascimb\`ene}}, \bibinfo {author}
  {\bibfnamefont {J.}~\bibnamefont {Dalibard}}, \ and\ \bibinfo {author}
  {\bibfnamefont {J.}~\bibnamefont {Beugnon}},\ }\bibfield  {title} {\enquote
  {\bibinfo {title} {Transmission of near-resonant light through a dense slab
  of cold atoms},}\ }\href {\doibase 10.1103/PhysRevA.96.053629} {\bibfield
  {journal} {\bibinfo  {journal} {Phys. Rev. A}\ }\textbf {\bibinfo {volume}
  {96}},\ \bibinfo {pages} {053629} (\bibinfo {year} {2017})}\BibitemShut
  {NoStop}%
\bibitem [{\citenamefont {Bettles}\ \emph {et~al.}(2018)\citenamefont
  {Bettles}, \citenamefont {Ilieva}, \citenamefont {Busche}, \citenamefont
  {Huillery}, \citenamefont {Ball}, \citenamefont {Spong},\ and\ \citenamefont
  {Adams}}]{Bettles18}%
  \BibitemOpen
  \bibfield  {author} {\bibinfo {author} {\bibfnamefont {R.~J.}\ \bibnamefont
  {Bettles}}, \bibinfo {author} {\bibfnamefont {T.}~\bibnamefont {Ilieva}},
  \bibinfo {author} {\bibfnamefont {H.}~\bibnamefont {Busche}}, \bibinfo
  {author} {\bibfnamefont {P.}~\bibnamefont {Huillery}}, \bibinfo {author}
  {\bibfnamefont {S.~W.}\ \bibnamefont {Ball}}, \bibinfo {author}
  {\bibfnamefont {N.~L.~R.}\ \bibnamefont {Spong}}, \ and\ \bibinfo {author}
  {\bibfnamefont {C.~S.}\ \bibnamefont {Adams}},\ }\bibfield  {title} {\enquote
  {\bibinfo {title} {Collective mode interferences in light-matter
  interactions},}\ }\href@noop {} {\  (\bibinfo {year} {2018})},\ \Eprint
  {http://arxiv.org/abs/1808.08415} {arXiv:1808.08415} \BibitemShut {NoStop}%
\bibitem [{\citenamefont {Morice}\ \emph {et~al.}(1995)\citenamefont {Morice},
  \citenamefont {Castin},\ and\ \citenamefont {Dalibard}}]{Morice1995a}%
  \BibitemOpen
  \bibfield  {author} {\bibinfo {author} {\bibfnamefont {Olivier}\ \bibnamefont
  {Morice}}, \bibinfo {author} {\bibfnamefont {Yvan}\ \bibnamefont {Castin}}, \
  and\ \bibinfo {author} {\bibfnamefont {Jean}\ \bibnamefont {Dalibard}},\
  }\bibfield  {title} {\enquote {\bibinfo {title} {Refractive index of a dilute
  bose gas},}\ }\href {\doibase 10.1103/PhysRevA.51.3896} {\bibfield  {journal}
  {\bibinfo  {journal} {Phys. Rev. A}\ }\textbf {\bibinfo {volume} {51}},\
  \bibinfo {pages} {3896--3901} (\bibinfo {year} {1995})}\BibitemShut {NoStop}%
\bibitem [{\citenamefont {Ruostekoski}\ and\ \citenamefont
  {Javanainen}(1997{\natexlab{a}})}]{Ruostekoski1997a}%
  \BibitemOpen
  \bibfield  {author} {\bibinfo {author} {\bibfnamefont {Janne}\ \bibnamefont
  {Ruostekoski}}\ and\ \bibinfo {author} {\bibfnamefont {Juha}\ \bibnamefont
  {Javanainen}},\ }\bibfield  {title} {\enquote {\bibinfo {title} {Quantum
  field theory of cooperative atom response: Low light intensity},}\ }\href
  {\doibase 10.1103/PhysRevA.55.513} {\bibfield  {journal} {\bibinfo  {journal}
  {Phys. Rev. A}\ }\textbf {\bibinfo {volume} {55}},\ \bibinfo {pages}
  {513--526} (\bibinfo {year} {1997}{\natexlab{a}})}\BibitemShut {NoStop}%
\bibitem [{\citenamefont {Javanainen}\ \emph {et~al.}(2014)\citenamefont
  {Javanainen}, \citenamefont {Ruostekoski}, \citenamefont {Li},\ and\
  \citenamefont {Yoo}}]{Javanainen2014a}%
  \BibitemOpen
  \bibfield  {author} {\bibinfo {author} {\bibfnamefont {Juha}\ \bibnamefont
  {Javanainen}}, \bibinfo {author} {\bibfnamefont {Janne}\ \bibnamefont
  {Ruostekoski}}, \bibinfo {author} {\bibfnamefont {Yi}~\bibnamefont {Li}}, \
  and\ \bibinfo {author} {\bibfnamefont {Sung-Mi}\ \bibnamefont {Yoo}},\
  }\bibfield  {title} {\enquote {\bibinfo {title} {Shifts of a resonance line
  in a dense atomic sample},}\ }\href {\doibase 10.1103/PhysRevLett.112.113603}
  {\bibfield  {journal} {\bibinfo  {journal} {Phys. Rev. Lett.}\ }\textbf
  {\bibinfo {volume} {112}},\ \bibinfo {pages} {113603} (\bibinfo {year}
  {2014})}\BibitemShut {NoStop}%
\bibitem [{\citenamefont {Javanainen}\ and\ \citenamefont
  {Ruostekoski}(2016)}]{JavanainenMFT}%
  \BibitemOpen
  \bibfield  {author} {\bibinfo {author} {\bibfnamefont {Juha}\ \bibnamefont
  {Javanainen}}\ and\ \bibinfo {author} {\bibfnamefont {Janne}\ \bibnamefont
  {Ruostekoski}},\ }\bibfield  {title} {\enquote {\bibinfo {title} {Light
  propagation beyond the mean-field theory of standard optics},}\ }\href
  {\doibase 10.1364/OE.24.000993} {\bibfield  {journal} {\bibinfo  {journal}
  {Opt. Express}\ }\textbf {\bibinfo {volume} {24}},\ \bibinfo {pages}
  {993--1001} (\bibinfo {year} {2016})}\BibitemShut {NoStop}%
\bibitem [{\citenamefont {Lemoult}\ \emph {et~al.}(2010)\citenamefont
  {Lemoult}, \citenamefont {Lerosey}, \citenamefont {de~Rosny},\ and\
  \citenamefont {Fink}}]{lemoult2010}%
  \BibitemOpen
  \bibfield  {author} {\bibinfo {author} {\bibfnamefont {Fabrice}\ \bibnamefont
  {Lemoult}}, \bibinfo {author} {\bibfnamefont {Geoffroy}\ \bibnamefont
  {Lerosey}}, \bibinfo {author} {\bibfnamefont {Julien}\ \bibnamefont
  {de~Rosny}}, \ and\ \bibinfo {author} {\bibfnamefont {Mathias}\ \bibnamefont
  {Fink}},\ }\bibfield  {title} {\enquote {\bibinfo {title} {Resonant
  metalenses for breaking the diffraction barrier},}\ }\href {\doibase
  10.1103/PhysRevLett.104.203901} {\bibfield  {journal} {\bibinfo  {journal}
  {Phys. Rev. Lett.}\ }\textbf {\bibinfo {volume} {104}},\ \bibinfo {pages}
  {203901} (\bibinfo {year} {2010})}\BibitemShut {NoStop}%
\bibitem [{\citenamefont {Fedotov}\ \emph {et~al.}(2010)\citenamefont
  {Fedotov}, \citenamefont {Papasimakis}, \citenamefont {Plum}, \citenamefont
  {Bitzer}, \citenamefont {Walther}, \citenamefont {Kuo}, \citenamefont
  {Tsai},\ and\ \citenamefont {Zheludev}}]{Fedotov2010}%
  \BibitemOpen
  \bibfield  {author} {\bibinfo {author} {\bibfnamefont {Vassili~A.}\
  \bibnamefont {Fedotov}}, \bibinfo {author} {\bibfnamefont {N.}~\bibnamefont
  {Papasimakis}}, \bibinfo {author} {\bibfnamefont {E.}~\bibnamefont {Plum}},
  \bibinfo {author} {\bibfnamefont {A.}~\bibnamefont {Bitzer}}, \bibinfo
  {author} {\bibfnamefont {M.}~\bibnamefont {Walther}}, \bibinfo {author}
  {\bibfnamefont {P.}~\bibnamefont {Kuo}}, \bibinfo {author} {\bibfnamefont
  {D~P}\ \bibnamefont {Tsai}}, \ and\ \bibinfo {author} {\bibfnamefont {N~I}\
  \bibnamefont {Zheludev}},\ }\bibfield  {title} {\enquote {\bibinfo {title}
  {{Spectral Collapse in Ensembles of Metamolecules}},}\ }\href {\doibase
  10.1103/PhysRevLett.104.223901} {\bibfield  {journal} {\bibinfo  {journal}
  {Phys. Rev. Lett.}\ }\textbf {\bibinfo {volume} {104}},\ \bibinfo {pages}
  {223901} (\bibinfo {year} {2010})}\BibitemShut {NoStop}%
\bibitem [{\citenamefont {Jenkins}\ and\ \citenamefont
  {Ruostekoski}(2012)}]{Jenkins2012a}%
  \BibitemOpen
  \bibfield  {author} {\bibinfo {author} {\bibfnamefont {Stewart~D.}\
  \bibnamefont {Jenkins}}\ and\ \bibinfo {author} {\bibfnamefont {Janne}\
  \bibnamefont {Ruostekoski}},\ }\bibfield  {title} {\enquote {\bibinfo {title}
  {Controlled manipulation of light by cooperative response of atoms in an
  optical lattice},}\ }\href {\doibase 10.1103/PhysRevA.86.031602} {\bibfield
  {journal} {\bibinfo  {journal} {Phys. Rev. A}\ }\textbf {\bibinfo {volume}
  {86}},\ \bibinfo {pages} {031602(R)} (\bibinfo {year} {2012})}\BibitemShut
  {NoStop}%
\bibitem [{\citenamefont {Perczel}\ \emph {et~al.}(2017)\citenamefont
  {Perczel}, \citenamefont {Borregaard}, \citenamefont {Chang}, \citenamefont
  {Pichler}, \citenamefont {Yelin}, \citenamefont {Zoller},\ and\ \citenamefont
  {Lukin}}]{Perczel}%
  \BibitemOpen
  \bibfield  {author} {\bibinfo {author} {\bibfnamefont {J.}~\bibnamefont
  {Perczel}}, \bibinfo {author} {\bibfnamefont {J.}~\bibnamefont {Borregaard}},
  \bibinfo {author} {\bibfnamefont {D.~E.}\ \bibnamefont {Chang}}, \bibinfo
  {author} {\bibfnamefont {H.}~\bibnamefont {Pichler}}, \bibinfo {author}
  {\bibfnamefont {S.~F.}\ \bibnamefont {Yelin}}, \bibinfo {author}
  {\bibfnamefont {P.}~\bibnamefont {Zoller}}, \ and\ \bibinfo {author}
  {\bibfnamefont {M.~D.}\ \bibnamefont {Lukin}},\ }\bibfield  {title} {\enquote
  {\bibinfo {title} {Photonic band structure of two-dimensional atomic
  lattices},}\ }\href {\doibase 10.1103/PhysRevA.96.063801} {\bibfield
  {journal} {\bibinfo  {journal} {Phys. Rev. A}\ }\textbf {\bibinfo {volume}
  {96}},\ \bibinfo {pages} {063801} (\bibinfo {year} {2017})}\BibitemShut
  {NoStop}%
\bibitem [{\citenamefont {Bettles}\ \emph {et~al.}(2017)\citenamefont
  {Bettles}, \citenamefont {Min\'a\ifmmode~\check{r}\else \v{r}\fi{}},
  \citenamefont {Adams}, \citenamefont {Lesanovsky},\ and\ \citenamefont
  {Olmos}}]{Bettles_17topo}%
  \BibitemOpen
  \bibfield  {author} {\bibinfo {author} {\bibfnamefont {R.~J.}\ \bibnamefont
  {Bettles}}, \bibinfo {author} {\bibfnamefont {J.}~\bibnamefont
  {Min\'a\ifmmode~\check{r}\else \v{r}\fi{}}}, \bibinfo {author} {\bibfnamefont
  {C.~S.}\ \bibnamefont {Adams}}, \bibinfo {author} {\bibfnamefont
  {I.}~\bibnamefont {Lesanovsky}}, \ and\ \bibinfo {author} {\bibfnamefont
  {B.}~\bibnamefont {Olmos}},\ }\bibfield  {title} {\enquote {\bibinfo {title}
  {Topological properties of a dense atomic lattice gas},}\ }\href {\doibase
  10.1103/PhysRevA.96.041603} {\bibfield  {journal} {\bibinfo  {journal} {Phys.
  Rev. A}\ }\textbf {\bibinfo {volume} {96}},\ \bibinfo {pages} {041603}
  (\bibinfo {year} {2017})}\BibitemShut {NoStop}%
\bibitem [{\citenamefont {Jenkins}\ and\ \citenamefont
  {Ruostekoski}(2013)}]{CAIT}%
  \BibitemOpen
  \bibfield  {author} {\bibinfo {author} {\bibfnamefont {Stewart~D.}\
  \bibnamefont {Jenkins}}\ and\ \bibinfo {author} {\bibfnamefont {Janne}\
  \bibnamefont {Ruostekoski}},\ }\bibfield  {title} {\enquote {\bibinfo {title}
  {Metamaterial transparency induced by cooperative electromagnetic
  interactions},}\ }\href {\doibase 10.1103/PhysRevLett.111.147401} {\bibfield
  {journal} {\bibinfo  {journal} {Phys. Rev. Lett.}\ }\textbf {\bibinfo
  {volume} {111}},\ \bibinfo {pages} {147401} (\bibinfo {year}
  {2013})}\BibitemShut {NoStop}%
\bibitem [{\citenamefont {Bettles}\ \emph {et~al.}(2015)\citenamefont
  {Bettles}, \citenamefont {Gardiner},\ and\ \citenamefont
  {Adams}}]{Bettles2015d}%
  \BibitemOpen
  \bibfield  {author} {\bibinfo {author} {\bibfnamefont {Robert~J.}\
  \bibnamefont {Bettles}}, \bibinfo {author} {\bibfnamefont {S.~A.}\
  \bibnamefont {Gardiner}}, \ and\ \bibinfo {author} {\bibfnamefont
  {Charles~S.}\ \bibnamefont {Adams}},\ }\bibfield  {title} {\enquote {\bibinfo
  {title} {{Cooperative ordering in lattices of interacting two-level
  dipoles}},}\ }\href {\doibase 10.1103/PhysRevA.92.063822} {\bibfield
  {journal} {\bibinfo  {journal} {Phys. Rev. A}\ }\textbf {\bibinfo {volume}
  {92}},\ \bibinfo {pages} {063822} (\bibinfo {year} {2015})}\BibitemShut
  {NoStop}%
\bibitem [{\citenamefont {Bettles}\ \emph {et~al.}(2016)\citenamefont
  {Bettles}, \citenamefont {Gardiner},\ and\ \citenamefont
  {Adams}}]{Bettles2016}%
  \BibitemOpen
  \bibfield  {author} {\bibinfo {author} {\bibfnamefont {Robert~J.}\
  \bibnamefont {Bettles}}, \bibinfo {author} {\bibfnamefont {S.~A.}\
  \bibnamefont {Gardiner}}, \ and\ \bibinfo {author} {\bibfnamefont
  {Charles~S.}\ \bibnamefont {Adams}},\ }\bibfield  {title} {\enquote {\bibinfo
  {title} {{Enhanced Optical Cross Section via Collective Coupling of Atomic
  Dipoles in a 2D Array}},}\ }\href {\doibase 10.1103/PhysRevLett.116.103602}
  {\bibfield  {journal} {\bibinfo  {journal} {Phys. Rev. Lett.}\ }\textbf
  {\bibinfo {volume} {116}},\ \bibinfo {pages} {103602} (\bibinfo {year}
  {2016})}\BibitemShut {NoStop}%
\bibitem [{\citenamefont {Facchinetti}\ \emph {et~al.}(2016)\citenamefont
  {Facchinetti}, \citenamefont {Jenkins},\ and\ \citenamefont
  {Ruostekoski}}]{Facchinetti16}%
  \BibitemOpen
  \bibfield  {author} {\bibinfo {author} {\bibfnamefont {G.}~\bibnamefont
  {Facchinetti}}, \bibinfo {author} {\bibfnamefont {S.~D.}\ \bibnamefont
  {Jenkins}}, \ and\ \bibinfo {author} {\bibfnamefont {J.}~\bibnamefont
  {Ruostekoski}},\ }\bibfield  {title} {\enquote {\bibinfo {title} {Storing
  light with subradiant correlations in arrays of atoms},}\ }\href {\doibase
  10.1103/PhysRevLett.117.243601} {\bibfield  {journal} {\bibinfo  {journal}
  {Phys. Rev. Lett.}\ }\textbf {\bibinfo {volume} {117}},\ \bibinfo {pages}
  {243601} (\bibinfo {year} {2016})}\BibitemShut {NoStop}%
\bibitem [{\citenamefont {Yang}\ \emph {et~al.}(2014)\citenamefont {Yang},
  \citenamefont {Kravchenko}, \citenamefont {Briggs},\ and\ \citenamefont
  {Valentine}}]{valentine2014}%
  \BibitemOpen
  \bibfield  {author} {\bibinfo {author} {\bibfnamefont {Yuanmu}\ \bibnamefont
  {Yang}}, \bibinfo {author} {\bibfnamefont {Ivan~I.}\ \bibnamefont
  {Kravchenko}}, \bibinfo {author} {\bibfnamefont {Dayrl~P.}\ \bibnamefont
  {Briggs}}, \ and\ \bibinfo {author} {\bibfnamefont {Jason}\ \bibnamefont
  {Valentine}},\ }\bibfield  {title} {\enquote {\bibinfo {title}
  {All-dielectric metasurface analogue of electromagnetically induced
  transparency},}\ }\href {\doibase 10.1038/ncomms6753} {\bibfield  {journal}
  {\bibinfo  {journal} {Nature Comms.}\ }\textbf {\bibinfo {volume} {5}},\
  \bibinfo {pages} {5753} (\bibinfo {year} {2014})}\BibitemShut {NoStop}%
\bibitem [{\citenamefont {Facchinetti}\ and\ \citenamefont
  {Ruostekoski}(2018)}]{Facchinetti18}%
  \BibitemOpen
  \bibfield  {author} {\bibinfo {author} {\bibfnamefont {G.}~\bibnamefont
  {Facchinetti}}\ and\ \bibinfo {author} {\bibfnamefont {J.}~\bibnamefont
  {Ruostekoski}},\ }\bibfield  {title} {\enquote {\bibinfo {title} {Interaction
  of light with planar lattices of atoms: Reflection, transmission, and
  cooperative magnetometry},}\ }\href {\doibase 10.1103/PhysRevA.97.023833}
  {\bibfield  {journal} {\bibinfo  {journal} {Phys. Rev. A}\ }\textbf {\bibinfo
  {volume} {97}},\ \bibinfo {pages} {023833} (\bibinfo {year}
  {2018})}\BibitemShut {NoStop}%
\bibitem [{\citenamefont {Shahmoon}\ \emph {et~al.}(2017)\citenamefont
  {Shahmoon}, \citenamefont {Wild}, \citenamefont {Lukin},\ and\ \citenamefont
  {Yelin}}]{Shahmoon}%
  \BibitemOpen
  \bibfield  {author} {\bibinfo {author} {\bibfnamefont {Ephraim}\ \bibnamefont
  {Shahmoon}}, \bibinfo {author} {\bibfnamefont {Dominik~S.}\ \bibnamefont
  {Wild}}, \bibinfo {author} {\bibfnamefont {Mikhail~D.}\ \bibnamefont
  {Lukin}}, \ and\ \bibinfo {author} {\bibfnamefont {Susanne~F.}\ \bibnamefont
  {Yelin}},\ }\bibfield  {title} {\enquote {\bibinfo {title} {Cooperative
  resonances in light scattering from two-dimensional atomic arrays},}\ }\href
  {\doibase 10.1103/PhysRevLett.118.113601} {\bibfield  {journal} {\bibinfo
  {journal} {Phys. Rev. Lett.}\ }\textbf {\bibinfo {volume} {118}},\ \bibinfo
  {pages} {113601} (\bibinfo {year} {2017})}\BibitemShut {NoStop}%
\bibitem [{\citenamefont {Plankensteiner}\ \emph {et~al.}(2017)\citenamefont
  {Plankensteiner}, \citenamefont {Sommer}, \citenamefont {Ritsch},\ and\
  \citenamefont {Genes}}]{Plankensteiner2017}%
  \BibitemOpen
  \bibfield  {author} {\bibinfo {author} {\bibfnamefont {David}\ \bibnamefont
  {Plankensteiner}}, \bibinfo {author} {\bibfnamefont {Christian}\ \bibnamefont
  {Sommer}}, \bibinfo {author} {\bibfnamefont {Helmut}\ \bibnamefont {Ritsch}},
  \ and\ \bibinfo {author} {\bibfnamefont {Claudiu}\ \bibnamefont {Genes}},\
  }\bibfield  {title} {\enquote {\bibinfo {title} {{Cavity Antiresonance
  Spectroscopy of Dipole Coupled Subradiant Arrays}},}\ }\href {\doibase
  10.1103/PhysRevLett.119.093601} {\bibfield  {journal} {\bibinfo  {journal}
  {Phys. Rev. Lett.}\ }\textbf {\bibinfo {volume} {119}},\ \bibinfo {pages}
  {093601} (\bibinfo {year} {2017})}\BibitemShut {NoStop}%
\bibitem [{\citenamefont {Asenjo-Garcia}\ \emph {et~al.}(2017)\citenamefont
  {Asenjo-Garcia}, \citenamefont {Moreno-Cardoner}, \citenamefont {Albrecht},
  \citenamefont {Kimble},\ and\ \citenamefont {Chang}}]{Asenjo-Garcia2017a}%
  \BibitemOpen
  \bibfield  {author} {\bibinfo {author} {\bibfnamefont {Ana}\ \bibnamefont
  {Asenjo-Garcia}}, \bibinfo {author} {\bibfnamefont {M.}~\bibnamefont
  {Moreno-Cardoner}}, \bibinfo {author} {\bibfnamefont {A.}~\bibnamefont
  {Albrecht}}, \bibinfo {author} {\bibfnamefont {H.~J.}\ \bibnamefont
  {Kimble}}, \ and\ \bibinfo {author} {\bibfnamefont {D.~E.}\ \bibnamefont
  {Chang}},\ }\bibfield  {title} {\enquote {\bibinfo {title} {{Exponential
  Improvement in Photon Storage Fidelities Using Subradiance and “Selective
  Radiance” in Atomic Arrays}},}\ }\href {\doibase 10.1103/PhysRevX.7.031024}
  {\bibfield  {journal} {\bibinfo  {journal} {Phys. Rev. X}\ }\textbf {\bibinfo
  {volume} {7}},\ \bibinfo {pages} {031024} (\bibinfo {year}
  {2017})}\BibitemShut {NoStop}%
\bibitem [{\citenamefont {Jen}(2017)}]{Jen17}%
  \BibitemOpen
  \bibfield  {author} {\bibinfo {author} {\bibfnamefont {H.~H.}\ \bibnamefont
  {Jen}},\ }\bibfield  {title} {\enquote {\bibinfo {title} {Phase-imprinted
  multiphoton subradiant states},}\ }\href {\doibase
  10.1103/PhysRevA.96.023814} {\bibfield  {journal} {\bibinfo  {journal} {Phys.
  Rev. A}\ }\textbf {\bibinfo {volume} {96}},\ \bibinfo {pages} {023814}
  (\bibinfo {year} {2017})}\BibitemShut {NoStop}%
\bibitem [{\citenamefont {Guimond}\ \emph {et~al.}(2019)\citenamefont
  {Guimond}, \citenamefont {Grankin}, \citenamefont {Vasilyev}, \citenamefont
  {Vermersch},\ and\ \citenamefont {Zoller}}]{Guimond2019}%
  \BibitemOpen
  \bibfield  {author} {\bibinfo {author} {\bibfnamefont {P.-O.}\ \bibnamefont
  {Guimond}}, \bibinfo {author} {\bibfnamefont {A.}~\bibnamefont {Grankin}},
  \bibinfo {author} {\bibfnamefont {D.~V.}\ \bibnamefont {Vasilyev}}, \bibinfo
  {author} {\bibfnamefont {B.}~\bibnamefont {Vermersch}}, \ and\ \bibinfo
  {author} {\bibfnamefont {P.}~\bibnamefont {Zoller}},\ }\bibfield  {title}
  {\enquote {\bibinfo {title} {Subradiant bell states in distant atomic
  arrays},}\ }\href {\doibase 10.1103/PhysRevLett.122.093601} {\bibfield
  {journal} {\bibinfo  {journal} {Phys. Rev. Lett.}\ }\textbf {\bibinfo
  {volume} {122}},\ \bibinfo {pages} {093601} (\bibinfo {year}
  {2019})}\BibitemShut {NoStop}%
\bibitem [{\citenamefont {Hebenstreit}\ \emph {et~al.}(2017)\citenamefont
  {Hebenstreit}, \citenamefont {Kraus}, \citenamefont {Ostermann},\ and\
  \citenamefont {Ritsch}}]{Ritsch_subr}%
  \BibitemOpen
  \bibfield  {author} {\bibinfo {author} {\bibfnamefont {Martin}\ \bibnamefont
  {Hebenstreit}}, \bibinfo {author} {\bibfnamefont {Barbara}\ \bibnamefont
  {Kraus}}, \bibinfo {author} {\bibfnamefont {Laurin}\ \bibnamefont
  {Ostermann}}, \ and\ \bibinfo {author} {\bibfnamefont {Helmut}\ \bibnamefont
  {Ritsch}},\ }\bibfield  {title} {\enquote {\bibinfo {title} {Subradiance via
  entanglement in atoms with several independent decay channels},}\ }\href
  {\doibase 10.1103/PhysRevLett.118.143602} {\bibfield  {journal} {\bibinfo
  {journal} {Phys. Rev. Lett.}\ }\textbf {\bibinfo {volume} {118}},\ \bibinfo
  {pages} {143602} (\bibinfo {year} {2017})}\BibitemShut {NoStop}%
\bibitem [{\citenamefont {Grankin}\ \emph {et~al.}(2018)\citenamefont
  {Grankin}, \citenamefont {Guimond}, \citenamefont {Vasilyev}, \citenamefont
  {Vermersch},\ and\ \citenamefont {Zoller}}]{Grankin18}%
  \BibitemOpen
  \bibfield  {author} {\bibinfo {author} {\bibfnamefont {A.}~\bibnamefont
  {Grankin}}, \bibinfo {author} {\bibfnamefont {P.~O.}\ \bibnamefont
  {Guimond}}, \bibinfo {author} {\bibfnamefont {D.~V.}\ \bibnamefont
  {Vasilyev}}, \bibinfo {author} {\bibfnamefont {B.}~\bibnamefont {Vermersch}},
  \ and\ \bibinfo {author} {\bibfnamefont {P.}~\bibnamefont {Zoller}},\
  }\bibfield  {title} {\enquote {\bibinfo {title} {Free-space photonic quantum
  link and chiral quantum optics},}\ }\href {\doibase
  10.1103/PhysRevA.98.043825} {\bibfield  {journal} {\bibinfo  {journal} {Phys.
  Rev. A}\ }\textbf {\bibinfo {volume} {98}},\ \bibinfo {pages} {043825}
  (\bibinfo {year} {2018})}\BibitemShut {NoStop}%
\bibitem [{\citenamefont {Olmos}\ \emph {et~al.}(2013)\citenamefont {Olmos},
  \citenamefont {Yu}, \citenamefont {Singh}, \citenamefont {Schreck},
  \citenamefont {Bongs},\ and\ \citenamefont {Lesanovsky}}]{Olmos13}%
  \BibitemOpen
  \bibfield  {author} {\bibinfo {author} {\bibfnamefont {B.}~\bibnamefont
  {Olmos}}, \bibinfo {author} {\bibfnamefont {D.}~\bibnamefont {Yu}}, \bibinfo
  {author} {\bibfnamefont {Y.}~\bibnamefont {Singh}}, \bibinfo {author}
  {\bibfnamefont {F.}~\bibnamefont {Schreck}}, \bibinfo {author} {\bibfnamefont
  {K.}~\bibnamefont {Bongs}}, \ and\ \bibinfo {author} {\bibfnamefont
  {I.}~\bibnamefont {Lesanovsky}},\ }\bibfield  {title} {\enquote {\bibinfo
  {title} {Long-range interacting many-body systems with alkaline-earth-metal
  atoms},}\ }\href {\doibase 10.1103/PhysRevLett.110.143602} {\bibfield
  {journal} {\bibinfo  {journal} {Phys. Rev. Lett.}\ }\textbf {\bibinfo
  {volume} {110}},\ \bibinfo {pages} {143602} (\bibinfo {year}
  {2013})}\BibitemShut {NoStop}%
\bibitem [{\citenamefont {Kr{\"{a}}mer}\ \emph {et~al.}(2016)\citenamefont
  {Kr{\"{a}}mer}, \citenamefont {Ostermann},\ and\ \citenamefont
  {Ritsch}}]{Kramer2016}%
  \BibitemOpen
  \bibfield  {author} {\bibinfo {author} {\bibfnamefont {Sebastian}\
  \bibnamefont {Kr{\"{a}}mer}}, \bibinfo {author} {\bibfnamefont {Laurin}\
  \bibnamefont {Ostermann}}, \ and\ \bibinfo {author} {\bibfnamefont {Helmut}\
  \bibnamefont {Ritsch}},\ }\bibfield  {title} {\enquote {\bibinfo {title}
  {{Optimized geometries for future generation optical lattice clocks}},}\
  }\href {\doibase 10.1209/0295-5075/114/14003} {\bibfield  {journal} {\bibinfo
   {journal} {Europhys. Lett.}\ }\textbf {\bibinfo {volume} {114}},\ \bibinfo
  {pages} {14003} (\bibinfo {year} {2016})}\BibitemShut {NoStop}%
\bibitem [{\citenamefont {Sutherland}\ and\ \citenamefont
  {Robicheaux}(2016)}]{Sutherland1D}%
  \BibitemOpen
  \bibfield  {author} {\bibinfo {author} {\bibfnamefont {R.~T.}\ \bibnamefont
  {Sutherland}}\ and\ \bibinfo {author} {\bibfnamefont {F.}~\bibnamefont
  {Robicheaux}},\ }\bibfield  {title} {\enquote {\bibinfo {title} {Collective
  dipole-dipole interactions in an atomic array},}\ }\href {\doibase
  10.1103/PhysRevA.94.013847} {\bibfield  {journal} {\bibinfo  {journal} {Phys.
  Rev. A}\ }\textbf {\bibinfo {volume} {94}},\ \bibinfo {pages} {013847}
  (\bibinfo {year} {2016})}\BibitemShut {NoStop}%
\bibitem [{\citenamefont {Yoo}\ and\ \citenamefont {Paik}(2016)}]{Yoo2016}%
  \BibitemOpen
  \bibfield  {author} {\bibinfo {author} {\bibfnamefont {Sung-Mi}\ \bibnamefont
  {Yoo}}\ and\ \bibinfo {author} {\bibfnamefont {Sun~Mok}\ \bibnamefont
  {Paik}},\ }\bibfield  {title} {\enquote {\bibinfo {title} {{Cooperative
  optical response of 2D dense lattices with strongly correlated dipoles}},}\
  }\href {\doibase 10.1364/OE.24.002156} {\bibfield  {journal} {\bibinfo
  {journal} {Opt. Express}\ }\textbf {\bibinfo {volume} {24}},\ \bibinfo
  {pages} {2156} (\bibinfo {year} {2016})}\BibitemShut {NoStop}%
\bibitem [{\citenamefont {Wang}\ \emph {et~al.}(2017)\citenamefont {Wang},
  \citenamefont {Zhao}, \citenamefont {Kan},\ and\ \citenamefont
  {Huang}}]{Wang2017}%
  \BibitemOpen
  \bibfield  {author} {\bibinfo {author} {\bibfnamefont {B~X}\ \bibnamefont
  {Wang}}, \bibinfo {author} {\bibfnamefont {C~Y}\ \bibnamefont {Zhao}},
  \bibinfo {author} {\bibfnamefont {Y~H}\ \bibnamefont {Kan}}, \ and\ \bibinfo
  {author} {\bibfnamefont {T~C}\ \bibnamefont {Huang}},\ }\bibfield  {title}
  {\enquote {\bibinfo {title} {{Design of metasurface polarizers based on
  two-dimensional cold atomic arrays}},}\ }\href {\doibase
  10.1364/OE.25.018760} {\bibfield  {journal} {\bibinfo  {journal} {Opt.
  Express}\ }\textbf {\bibinfo {volume} {25}},\ \bibinfo {pages} {18760}
  (\bibinfo {year} {2017})}\BibitemShut {NoStop}%
\bibitem [{\citenamefont {Jenkins}\ \emph {et~al.}(2017)\citenamefont
  {Jenkins}, \citenamefont {Ruostekoski}, \citenamefont {Papasimakis},
  \citenamefont {Savo},\ and\ \citenamefont {Zheludev}}]{Jenkins17}%
  \BibitemOpen
  \bibfield  {author} {\bibinfo {author} {\bibfnamefont {Stewart~D.}\
  \bibnamefont {Jenkins}}, \bibinfo {author} {\bibfnamefont {Janne}\
  \bibnamefont {Ruostekoski}}, \bibinfo {author} {\bibfnamefont {Nikitas}\
  \bibnamefont {Papasimakis}}, \bibinfo {author} {\bibfnamefont {Salvatore}\
  \bibnamefont {Savo}}, \ and\ \bibinfo {author} {\bibfnamefont {Nikolay~I.}\
  \bibnamefont {Zheludev}},\ }\bibfield  {title} {\enquote {\bibinfo {title}
  {Many-body subradiant excitations in metamaterial arrays: Experiment and
  theory},}\ }\href {\doibase 10.1103/PhysRevLett.119.053901} {\bibfield
  {journal} {\bibinfo  {journal} {Phys. Rev. Lett.}\ }\textbf {\bibinfo
  {volume} {119}},\ \bibinfo {pages} {053901} (\bibinfo {year}
  {2017})}\BibitemShut {NoStop}%
\bibitem [{\citenamefont {Yoo}(2018)}]{Yoo18}%
  \BibitemOpen
  \bibfield  {author} {\bibinfo {author} {\bibfnamefont {Sung-Mi}\ \bibnamefont
  {Yoo}},\ }\bibfield  {title} {\enquote {\bibinfo {title} {Strongly coupled
  cold atoms in bilayer dense lattices},}\ }\href
  {http://stacks.iop.org/1367-2630/20/i=8/a=083012} {\bibfield  {journal}
  {\bibinfo  {journal} {New Journal of Physics}\ }\textbf {\bibinfo {volume}
  {20}},\ \bibinfo {pages} {083012} (\bibinfo {year} {2018})}\BibitemShut
  {NoStop}%
\bibitem [{\citenamefont {Zeytino\v{g}lu}\ and\ \citenamefont
  {\.{I}mamo\v{g}lu}(2018)}]{Imamoglu2018b}%
  \BibitemOpen
  \bibfield  {author} {\bibinfo {author} {\bibfnamefont {Sina}\ \bibnamefont
  {Zeytino\v{g}lu}}\ and\ \bibinfo {author} {\bibfnamefont {Atac}\ \bibnamefont
  {\.{I}mamo\v{g}lu}},\ }\bibfield  {title} {\enquote {\bibinfo {title}
  {{Interaction-induced photon blockade using an atomically thin mirror
  embedded in a microcavity}},}\ }\href {\doibase 10.1103/PhysRevA.98.051801}
  {\bibfield  {journal} {\bibinfo  {journal} {Phys. Rev. A}\ }\textbf {\bibinfo
  {volume} {98}},\ \bibinfo {pages} {051801(R)} (\bibinfo {year}
  {2018})}\BibitemShut {NoStop}%
\bibitem [{\citenamefont {Manzoni}\ \emph {et~al.}(2018)\citenamefont
  {Manzoni}, \citenamefont {Moreno-Cardoner}, \citenamefont {Asenjo-Garcia},
  \citenamefont {Porto}, \citenamefont {Gorshkov},\ and\ \citenamefont
  {Chang}}]{Manzoni_2018}%
  \BibitemOpen
  \bibfield  {author} {\bibinfo {author} {\bibfnamefont {M~T}\ \bibnamefont
  {Manzoni}}, \bibinfo {author} {\bibfnamefont {M}~\bibnamefont
  {Moreno-Cardoner}}, \bibinfo {author} {\bibfnamefont {A}~\bibnamefont
  {Asenjo-Garcia}}, \bibinfo {author} {\bibfnamefont {J~V}\ \bibnamefont
  {Porto}}, \bibinfo {author} {\bibfnamefont {A~V}\ \bibnamefont {Gorshkov}}, \
  and\ \bibinfo {author} {\bibfnamefont {D~E}\ \bibnamefont {Chang}},\
  }\bibfield  {title} {\enquote {\bibinfo {title} {Optimization of photon
  storage fidelity in ordered atomic arrays},}\ }\href {\doibase
  10.1088/1367-2630/aadb74} {\bibfield  {journal} {\bibinfo  {journal} {New
  Journal of Physics}\ }\textbf {\bibinfo {volume} {20}},\ \bibinfo {pages}
  {083048} (\bibinfo {year} {2018})}\BibitemShut {NoStop}%
\bibitem [{\citenamefont {Mkhitaryan}\ \emph {et~al.}(2018)\citenamefont
  {Mkhitaryan}, \citenamefont {Meng}, \citenamefont {Marini},\ and\
  \citenamefont {de~Abajo}}]{Mkhitaryan2018}%
  \BibitemOpen
  \bibfield  {author} {\bibinfo {author} {\bibfnamefont {Vahagn}\ \bibnamefont
  {Mkhitaryan}}, \bibinfo {author} {\bibfnamefont {Lijun}\ \bibnamefont
  {Meng}}, \bibinfo {author} {\bibfnamefont {Andrea}\ \bibnamefont {Marini}}, \
  and\ \bibinfo {author} {\bibfnamefont {F.~Javier~Garc{\'{i}}a}\ \bibnamefont
  {de~Abajo}},\ }\bibfield  {title} {\enquote {\bibinfo {title} {{Lasing and
  Amplification from Two-Dimensional Atom Arrays}},}\ }\href {\doibase
  10.1103/PhysRevLett.121.163602} {\bibfield  {journal} {\bibinfo  {journal}
  {Phys. Rev. Lett.}\ }\textbf {\bibinfo {volume} {121}},\ \bibinfo {pages}
  {163602} (\bibinfo {year} {2018})}\BibitemShut {NoStop}%
\bibitem [{\citenamefont {Jen}\ \emph {et~al.}(2018)\citenamefont {Jen},
  \citenamefont {Chang},\ and\ \citenamefont {Chen}}]{Jen18}%
  \BibitemOpen
  \bibfield  {author} {\bibinfo {author} {\bibfnamefont {H.~H.}\ \bibnamefont
  {Jen}}, \bibinfo {author} {\bibfnamefont {M.~S.}\ \bibnamefont {Chang}}, \
  and\ \bibinfo {author} {\bibfnamefont {Y.~C.}\ \bibnamefont {Chen}},\
  }\bibfield  {title} {\enquote {\bibinfo {title} {Cooperative light scattering
  from helical-phase-imprinted atomic rings},}\ }\href {\doibase
  10.1038/s41598-018-27888-y} {\bibfield  {journal} {\bibinfo  {journal}
  {Scientific Reports}\ }\textbf {\bibinfo {volume} {8}},\ \bibinfo {pages}
  {9570} (\bibinfo {year} {2018})}\BibitemShut {NoStop}%
\bibitem [{\citenamefont {Zhang}\ and\ \citenamefont
  {M\o{}lmer}(2019)}]{Zhang2018}%
  \BibitemOpen
  \bibfield  {author} {\bibinfo {author} {\bibfnamefont {Yu-Xiang}\
  \bibnamefont {Zhang}}\ and\ \bibinfo {author} {\bibfnamefont {Klaus}\
  \bibnamefont {M\o{}lmer}},\ }\bibfield  {title} {\enquote {\bibinfo {title}
  {Theory of subradiant states of a one-dimensional two-level atom chain},}\
  }\href {\doibase 10.1103/PhysRevLett.122.203605} {\bibfield  {journal}
  {\bibinfo  {journal} {Phys. Rev. Lett.}\ }\textbf {\bibinfo {volume} {122}},\
  \bibinfo {pages} {203605} (\bibinfo {year} {2019})}\BibitemShut {NoStop}%
\bibitem [{\citenamefont {Henriet}\ \emph {et~al.}(2019)\citenamefont
  {Henriet}, \citenamefont {Douglas}, \citenamefont {Chang},\ and\
  \citenamefont {Albrecht}}]{Henriet2018}%
  \BibitemOpen
  \bibfield  {author} {\bibinfo {author} {\bibfnamefont {Lo\"{\i}c}\
  \bibnamefont {Henriet}}, \bibinfo {author} {\bibfnamefont {James~S.}\
  \bibnamefont {Douglas}}, \bibinfo {author} {\bibfnamefont {Darrick~E.}\
  \bibnamefont {Chang}}, \ and\ \bibinfo {author} {\bibfnamefont {Andreas}\
  \bibnamefont {Albrecht}},\ }\bibfield  {title} {\enquote {\bibinfo {title}
  {Critical open-system dynamics in a one-dimensional optical-lattice clock},}\
  }\href {\doibase 10.1103/PhysRevA.99.023802} {\bibfield  {journal} {\bibinfo
  {journal} {Phys. Rev. A}\ }\textbf {\bibinfo {volume} {99}},\ \bibinfo
  {pages} {023802} (\bibinfo {year} {2019})}\BibitemShut {NoStop}%
\bibitem [{\citenamefont {Parmee}\ and\ \citenamefont
  {Cooper}(2018)}]{Parmee2018}%
  \BibitemOpen
  \bibfield  {author} {\bibinfo {author} {\bibfnamefont {C.~D.}\ \bibnamefont
  {Parmee}}\ and\ \bibinfo {author} {\bibfnamefont {N.~R.}\ \bibnamefont
  {Cooper}},\ }\bibfield  {title} {\enquote {\bibinfo {title} {Phases of driven
  two-level systems with nonlocal dissipation},}\ }\href {\doibase
  10.1103/PhysRevA.97.053616} {\bibfield  {journal} {\bibinfo  {journal} {Phys.
  Rev. A}\ }\textbf {\bibinfo {volume} {97}},\ \bibinfo {pages} {053616}
  (\bibinfo {year} {2018})}\BibitemShut {NoStop}%
\bibitem [{\citenamefont {Plankensteiner}\ \emph {et~al.}(2019)\citenamefont
  {Plankensteiner}, \citenamefont {Sommer}, \citenamefont {Reitz},
  \citenamefont {Ritsch},\ and\ \citenamefont {Genes}}]{Plankensteiner19}%
  \BibitemOpen
  \bibfield  {author} {\bibinfo {author} {\bibfnamefont {D.}~\bibnamefont
  {Plankensteiner}}, \bibinfo {author} {\bibfnamefont {C.}~\bibnamefont
  {Sommer}}, \bibinfo {author} {\bibfnamefont {M.}~\bibnamefont {Reitz}},
  \bibinfo {author} {\bibfnamefont {H.}~\bibnamefont {Ritsch}}, \ and\ \bibinfo
  {author} {\bibfnamefont {C.}~\bibnamefont {Genes}},\ }\bibfield  {title}
  {\enquote {\bibinfo {title} {Enhanced collective purcell effect of coupled
  quantum emitter systems},}\ }\href {\doibase 10.1103/PhysRevA.99.043843}
  {\bibfield  {journal} {\bibinfo  {journal} {Phys. Rev. A}\ }\textbf {\bibinfo
  {volume} {99}},\ \bibinfo {pages} {043843} (\bibinfo {year}
  {2019})}\BibitemShut {NoStop}%
\bibitem [{\citenamefont {Javanainen}\ and\ \citenamefont
  {Rajapakse}(2019)}]{Javanainen19}%
  \BibitemOpen
  \bibfield  {author} {\bibinfo {author} {\bibfnamefont {Juha}\ \bibnamefont
  {Javanainen}}\ and\ \bibinfo {author} {\bibfnamefont {Renuka}\ \bibnamefont
  {Rajapakse}},\ }\bibfield  {title} {\enquote {\bibinfo {title} {Light
  propagation in systems involving two-dimensional atomic lattices},}\ }\href
  {\doibase 10.1103/PhysRevA.100.013616} {\bibfield  {journal} {\bibinfo
  {journal} {Phys. Rev. A}\ }\textbf {\bibinfo {volume} {100}},\ \bibinfo
  {pages} {013616} (\bibinfo {year} {2019})}\BibitemShut {NoStop}%
\bibitem [{\citenamefont {Choi}\ \emph {et~al.}(2008)\citenamefont {Choi},
  \citenamefont {Deng}, \citenamefont {Laurat},\ and\ \citenamefont
  {Kimble}}]{ChoiEtAlNature2008}%
  \BibitemOpen
  \bibfield  {author} {\bibinfo {author} {\bibfnamefont {K.~S.}\ \bibnamefont
  {Choi}}, \bibinfo {author} {\bibfnamefont {H.}~\bibnamefont {Deng}}, \bibinfo
  {author} {\bibfnamefont {J.}~\bibnamefont {Laurat}}, \ and\ \bibinfo {author}
  {\bibfnamefont {H.~J.}\ \bibnamefont {Kimble}},\ }\bibfield  {title}
  {\enquote {\bibinfo {title} {Mapping photonic entanglement into and out of a
  quantum memory},}\ }\href {https://doi.org/10.1038/nature06670} {\bibfield
  {journal} {\bibinfo  {journal} {Nature}\ }\textbf {\bibinfo {volume} {452}},\
  \bibinfo {pages} {67} (\bibinfo {year} {2008})}\BibitemShut {NoStop}%
\bibitem [{\citenamefont {Duan}\ and\ \citenamefont {Monroe}(2010)}]{Duan10}%
  \BibitemOpen
  \bibfield  {author} {\bibinfo {author} {\bibfnamefont {L.-M.}\ \bibnamefont
  {Duan}}\ and\ \bibinfo {author} {\bibfnamefont {C.}~\bibnamefont {Monroe}},\
  }\bibfield  {title} {\enquote {\bibinfo {title} {Colloquium: Quantum networks
  with trapped ions},}\ }\href {\doibase 10.1103/RevModPhys.82.1209} {\bibfield
   {journal} {\bibinfo  {journal} {Rev. Mod. Phys.}\ }\textbf {\bibinfo
  {volume} {82}},\ \bibinfo {pages} {1209--1224} (\bibinfo {year}
  {2010})}\BibitemShut {NoStop}%
\bibitem [{\citenamefont {Nickerson}\ \emph {et~al.}(2014)\citenamefont
  {Nickerson}, \citenamefont {Fitzsimons},\ and\ \citenamefont
  {Benjamin}}]{Nickerson14}%
  \BibitemOpen
  \bibfield  {author} {\bibinfo {author} {\bibfnamefont {Naomi~H.}\
  \bibnamefont {Nickerson}}, \bibinfo {author} {\bibfnamefont {Joseph~F.}\
  \bibnamefont {Fitzsimons}}, \ and\ \bibinfo {author} {\bibfnamefont
  {Simon~C.}\ \bibnamefont {Benjamin}},\ }\bibfield  {title} {\enquote
  {\bibinfo {title} {Freely scalable quantum technologies using cells of
  5-to-50 qubits with very lossy and noisy photonic links},}\ }\href {\doibase
  10.1103/PhysRevX.4.041041} {\bibfield  {journal} {\bibinfo  {journal} {Phys.
  Rev. X}\ }\textbf {\bibinfo {volume} {4}},\ \bibinfo {pages} {041041}
  (\bibinfo {year} {2014})}\BibitemShut {NoStop}%
\bibitem [{\citenamefont {Monroe}\ \emph {et~al.}(2014)\citenamefont {Monroe},
  \citenamefont {Raussendorf}, \citenamefont {Ruthven}, \citenamefont {Brown},
  \citenamefont {Maunz}, \citenamefont {Duan},\ and\ \citenamefont
  {Kim}}]{Monroe14}%
  \BibitemOpen
  \bibfield  {author} {\bibinfo {author} {\bibfnamefont {C.}~\bibnamefont
  {Monroe}}, \bibinfo {author} {\bibfnamefont {R.}~\bibnamefont {Raussendorf}},
  \bibinfo {author} {\bibfnamefont {A.}~\bibnamefont {Ruthven}}, \bibinfo
  {author} {\bibfnamefont {K.~R.}\ \bibnamefont {Brown}}, \bibinfo {author}
  {\bibfnamefont {P.}~\bibnamefont {Maunz}}, \bibinfo {author} {\bibfnamefont
  {L.-M.}\ \bibnamefont {Duan}}, \ and\ \bibinfo {author} {\bibfnamefont
  {J.}~\bibnamefont {Kim}},\ }\bibfield  {title} {\enquote {\bibinfo {title}
  {Large-scale modular quantum-computer architecture with atomic memory and
  photonic interconnects},}\ }\href {\doibase 10.1103/PhysRevA.89.022317}
  {\bibfield  {journal} {\bibinfo  {journal} {Phys. Rev. A}\ }\textbf {\bibinfo
  {volume} {89}},\ \bibinfo {pages} {022317} (\bibinfo {year}
  {2014})}\BibitemShut {NoStop}%
\bibitem [{\citenamefont {Dicke}(1954)}]{Dicke54}%
  \BibitemOpen
  \bibfield  {author} {\bibinfo {author} {\bibfnamefont {R.~H.}\ \bibnamefont
  {Dicke}},\ }\bibfield  {title} {\enquote {\bibinfo {title} {Coherence in
  spontaneous radiation processes},}\ }\href {\doibase 10.1103/PhysRev.93.99}
  {\bibfield  {journal} {\bibinfo  {journal} {Phys. Rev.}\ }\textbf {\bibinfo
  {volume} {93}},\ \bibinfo {pages} {99--110} (\bibinfo {year}
  {1954})}\BibitemShut {NoStop}%
\bibitem [{\citenamefont {DeVoe}\ and\ \citenamefont {Brewer}(1996)}]{DeVoe}%
  \BibitemOpen
  \bibfield  {author} {\bibinfo {author} {\bibfnamefont {R.~G.}\ \bibnamefont
  {DeVoe}}\ and\ \bibinfo {author} {\bibfnamefont {R.~G.}\ \bibnamefont
  {Brewer}},\ }\bibfield  {title} {\enquote {\bibinfo {title} {Observation of
  superradiant and subradiant spontaneous emission of two trapped ions},}\
  }\href {\doibase 10.1103/PhysRevLett.76.2049} {\bibfield  {journal} {\bibinfo
   {journal} {Phys. Rev. Lett.}\ }\textbf {\bibinfo {volume} {76}},\ \bibinfo
  {pages} {2049--2052} (\bibinfo {year} {1996})}\BibitemShut {NoStop}%
\bibitem [{\citenamefont {Hettich}\ \emph {et~al.}(2002)\citenamefont
  {Hettich}, \citenamefont {Schmitt}, \citenamefont {Zitzmann}, \citenamefont
  {K\"ohn}, \citenamefont {Gerhardt},\ and\ \citenamefont
  {Sandoghdar}}]{Hettich}%
  \BibitemOpen
  \bibfield  {author} {\bibinfo {author} {\bibfnamefont {C.}~\bibnamefont
  {Hettich}}, \bibinfo {author} {\bibfnamefont {C.}~\bibnamefont {Schmitt}},
  \bibinfo {author} {\bibfnamefont {J.}~\bibnamefont {Zitzmann}}, \bibinfo
  {author} {\bibfnamefont {S.}~\bibnamefont {K\"ohn}}, \bibinfo {author}
  {\bibfnamefont {I.}~\bibnamefont {Gerhardt}}, \ and\ \bibinfo {author}
  {\bibfnamefont {V.}~\bibnamefont {Sandoghdar}},\ }\bibfield  {title}
  {\enquote {\bibinfo {title} {Nanometer resolution and coherent optical dipole
  coupling of two individual molecules},}\ }\href {\doibase
  10.1126/science.1075606} {\bibfield  {journal} {\bibinfo  {journal}
  {Science}\ }\textbf {\bibinfo {volume} {298}},\ \bibinfo {pages} {385--389}
  (\bibinfo {year} {2002})}\BibitemShut {NoStop}%
\bibitem [{\citenamefont {McGuyer}\ \emph {et~al.}(2015)\citenamefont
  {McGuyer}, \citenamefont {McDonald}, \citenamefont {Iwata}, \citenamefont
  {Tarallo}, \citenamefont {Skomorowski}, \citenamefont {Moszynski},\ and\
  \citenamefont {Zelevinsky}}]{McGuyer}%
  \BibitemOpen
  \bibfield  {author} {\bibinfo {author} {\bibfnamefont {B.~H.}\ \bibnamefont
  {McGuyer}}, \bibinfo {author} {\bibfnamefont {M.}~\bibnamefont {McDonald}},
  \bibinfo {author} {\bibfnamefont {G.~Z.}\ \bibnamefont {Iwata}}, \bibinfo
  {author} {\bibfnamefont {M.~G.}\ \bibnamefont {Tarallo}}, \bibinfo {author}
  {\bibfnamefont {W.}~\bibnamefont {Skomorowski}}, \bibinfo {author}
  {\bibfnamefont {R.}~\bibnamefont {Moszynski}}, \ and\ \bibinfo {author}
  {\bibfnamefont {T.}~\bibnamefont {Zelevinsky}},\ }\bibfield  {title}
  {\enquote {\bibinfo {title} {Precise study of asymptotic physics with
  subradiant ultracold molecules},}\ }\href {\doibase 10.1038/NPHYS3182}
  {\bibfield  {journal} {\bibinfo  {journal} {Nat. Phys.}\ }\textbf {\bibinfo
  {volume} {11}},\ \bibinfo {pages} {32--36} (\bibinfo {year}
  {2015})}\BibitemShut {NoStop}%
\bibitem [{\citenamefont {Takasu}\ \emph {et~al.}(2012)\citenamefont {Takasu},
  \citenamefont {Saito}, \citenamefont {Takahashi}, \citenamefont {Borkowski},
  \citenamefont {Ciury\l{}o},\ and\ \citenamefont {Julienne}}]{Takasu}%
  \BibitemOpen
  \bibfield  {author} {\bibinfo {author} {\bibfnamefont {Yosuke}\ \bibnamefont
  {Takasu}}, \bibinfo {author} {\bibfnamefont {Yutaka}\ \bibnamefont {Saito}},
  \bibinfo {author} {\bibfnamefont {Yoshiro}\ \bibnamefont {Takahashi}},
  \bibinfo {author} {\bibfnamefont {Mateusz}\ \bibnamefont {Borkowski}},
  \bibinfo {author} {\bibfnamefont {Roman}\ \bibnamefont {Ciury\l{}o}}, \ and\
  \bibinfo {author} {\bibfnamefont {Paul~S.}\ \bibnamefont {Julienne}},\
  }\bibfield  {title} {\enquote {\bibinfo {title} {Controlled production of
  subradiant states of a diatomic molecule in an optical lattice},}\ }\href
  {\doibase 10.1103/PhysRevLett.108.173002} {\bibfield  {journal} {\bibinfo
  {journal} {Phys. Rev. Lett.}\ }\textbf {\bibinfo {volume} {108}},\ \bibinfo
  {pages} {173002} (\bibinfo {year} {2012})}\BibitemShut {NoStop}%
\bibitem [{\citenamefont {Lovera}\ \emph {et~al.}(2013)\citenamefont {Lovera},
  \citenamefont {Gallinet}, \citenamefont {Nordlander},\ and\ \citenamefont
  {Martin}}]{Lovera}%
  \BibitemOpen
  \bibfield  {author} {\bibinfo {author} {\bibfnamefont {Andrea}\ \bibnamefont
  {Lovera}}, \bibinfo {author} {\bibfnamefont {Benjamin}\ \bibnamefont
  {Gallinet}}, \bibinfo {author} {\bibfnamefont {Peter}\ \bibnamefont
  {Nordlander}}, \ and\ \bibinfo {author} {\bibfnamefont {Olivier~J.F.}\
  \bibnamefont {Martin}},\ }\bibfield  {title} {\enquote {\bibinfo {title}
  {Mechanisms of fano resonances in coupled plasmonic systems},}\ }\href
  {\doibase 10.1021/nn401175j} {\bibfield  {journal} {\bibinfo  {journal} {ACS
  Nano}\ }\textbf {\bibinfo {volume} {7}},\ \bibinfo {pages} {4527--4536}
  (\bibinfo {year} {2013})}\BibitemShut {NoStop}%
\bibitem [{\citenamefont {Frimmer}\ \emph {et~al.}({2012})\citenamefont
  {Frimmer}, \citenamefont {Coenen},\ and\ \citenamefont
  {Koenderink}}]{Frimmer}%
  \BibitemOpen
  \bibfield  {author} {\bibinfo {author} {\bibfnamefont {Martin}\ \bibnamefont
  {Frimmer}}, \bibinfo {author} {\bibfnamefont {Toon}\ \bibnamefont {Coenen}},
  \ and\ \bibinfo {author} {\bibfnamefont {A.~Femius}\ \bibnamefont
  {Koenderink}},\ }\bibfield  {title} {\enquote {\bibinfo {title} {{Signature
  of a Fano Resonance in a Plasmonic Metamolecule's Local Density of Optical
  States}},}\ }\href {\doibase {10.1103/PhysRevLett.108.077404}} {\bibfield
  {journal} {\bibinfo  {journal} {{Phys. Rev. Lett.}}\ }\textbf {\bibinfo
  {volume} {{108}}},\ \bibinfo {pages} {{077404}} (\bibinfo {year}
  {{2012}})}\BibitemShut {NoStop}%
\bibitem [{\citenamefont {Willingham}\ and\ \citenamefont
  {Link}(2011)}]{Willingham11}%
  \BibitemOpen
  \bibfield  {author} {\bibinfo {author} {\bibfnamefont {Britain}\ \bibnamefont
  {Willingham}}\ and\ \bibinfo {author} {\bibfnamefont {Stephan}\ \bibnamefont
  {Link}},\ }\bibfield  {title} {\enquote {\bibinfo {title} {Energy transport
  in metal nanoparticle chains via sub-radiant plasmon modes},}\ }\href
  {\doibase 10.1364/OE.19.006450} {\bibfield  {journal} {\bibinfo  {journal}
  {Opt. Express}\ }\textbf {\bibinfo {volume} {19}},\ \bibinfo {pages}
  {6450--6461} (\bibinfo {year} {2011})}\BibitemShut {NoStop}%
\bibitem [{\citenamefont {Giusteri}\ \emph {et~al.}(2015)\citenamefont
  {Giusteri}, \citenamefont {Mattiotti},\ and\ \citenamefont
  {Celardo}}]{Giusteri15}%
  \BibitemOpen
  \bibfield  {author} {\bibinfo {author} {\bibfnamefont {Giulio~G.}\
  \bibnamefont {Giusteri}}, \bibinfo {author} {\bibfnamefont {Francesco}\
  \bibnamefont {Mattiotti}}, \ and\ \bibinfo {author} {\bibfnamefont {G.~Luca}\
  \bibnamefont {Celardo}},\ }\bibfield  {title} {\enquote {\bibinfo {title}
  {Non-hermitian hamiltonian approach to quantum transport in disordered
  networks with sinks: Validity and effectiveness},}\ }\href {\doibase
  10.1103/PhysRevB.91.094301} {\bibfield  {journal} {\bibinfo  {journal} {Phys.
  Rev. B}\ }\textbf {\bibinfo {volume} {91}},\ \bibinfo {pages} {094301}
  (\bibinfo {year} {2015})}\BibitemShut {NoStop}%
\bibitem [{\citenamefont {Leggio}\ \emph {et~al.}(2015)\citenamefont {Leggio},
  \citenamefont {Messina},\ and\ \citenamefont {Antezza}}]{Leggio_2015}%
  \BibitemOpen
  \bibfield  {author} {\bibinfo {author} {\bibfnamefont {B.}~\bibnamefont
  {Leggio}}, \bibinfo {author} {\bibfnamefont {R.}~\bibnamefont {Messina}}, \
  and\ \bibinfo {author} {\bibfnamefont {M.}~\bibnamefont {Antezza}},\
  }\bibfield  {title} {\enquote {\bibinfo {title} {Thermally activated nonlocal
  amplification in quantum energy transport},}\ }\href {\doibase
  10.1209/0295-5075/110/40002} {\bibfield  {journal} {\bibinfo  {journal}
  {{EPL} (Europhysics Letters)}\ }\textbf {\bibinfo {volume} {110}},\ \bibinfo
  {pages} {40002} (\bibinfo {year} {2015})}\BibitemShut {NoStop}%
\bibitem [{\citenamefont {Doyeux}\ \emph {et~al.}(2017)\citenamefont {Doyeux},
  \citenamefont {Messina}, \citenamefont {Leggio},\ and\ \citenamefont
  {Antezza}}]{Doyeux17}%
  \BibitemOpen
  \bibfield  {author} {\bibinfo {author} {\bibfnamefont {Pierre}\ \bibnamefont
  {Doyeux}}, \bibinfo {author} {\bibfnamefont {Riccardo}\ \bibnamefont
  {Messina}}, \bibinfo {author} {\bibfnamefont {Bruno}\ \bibnamefont {Leggio}},
  \ and\ \bibinfo {author} {\bibfnamefont {Mauro}\ \bibnamefont {Antezza}},\
  }\bibfield  {title} {\enquote {\bibinfo {title} {Excitation injector in an
  atomic chain: Long-range transport and efficiency amplification},}\ }\href
  {\doibase 10.1103/PhysRevA.95.012138} {\bibfield  {journal} {\bibinfo
  {journal} {Phys. Rev. A}\ }\textbf {\bibinfo {volume} {95}},\ \bibinfo
  {pages} {012138} (\bibinfo {year} {2017})}\BibitemShut {NoStop}%
\bibitem [{\citenamefont {Jen}(2019)}]{Jen_2019}%
  \BibitemOpen
  \bibfield  {author} {\bibinfo {author} {\bibfnamefont {H~H}\ \bibnamefont
  {Jen}},\ }\bibfield  {title} {\enquote {\bibinfo {title} {Selective transport
  of atomic excitations in a driven chiral-coupled atomic chain},}\ }\href
  {\doibase 10.1088/1361-6455/ab04c1} {\bibfield  {journal} {\bibinfo
  {journal} {Journal of Physics B: Atomic, Molecular and Optical Physics}\
  }\textbf {\bibinfo {volume} {52}},\ \bibinfo {pages} {065502} (\bibinfo
  {year} {2019})}\BibitemShut {NoStop}%
\bibitem [{\citenamefont {Moreno-Cardoner}\ \emph {et~al.}(2019)\citenamefont
  {Moreno-Cardoner}, \citenamefont {Plankensteiner}, \citenamefont {Ostermann},
  \citenamefont {Chang},\ and\ \citenamefont {Ritsch}}]{Cardoner19}%
  \BibitemOpen
  \bibfield  {author} {\bibinfo {author} {\bibfnamefont {Maria}\ \bibnamefont
  {Moreno-Cardoner}}, \bibinfo {author} {\bibfnamefont {David}\ \bibnamefont
  {Plankensteiner}}, \bibinfo {author} {\bibfnamefont {Laurin}\ \bibnamefont
  {Ostermann}}, \bibinfo {author} {\bibfnamefont {Darrick~E.}\ \bibnamefont
  {Chang}}, \ and\ \bibinfo {author} {\bibfnamefont {Helmut}\ \bibnamefont
  {Ritsch}},\ }\bibfield  {title} {\enquote {\bibinfo {title}
  {Subradiance-enhanced excitation transfer between dipole-coupled nanorings of
  quantum emitters},}\ }\href {\doibase 10.1103/PhysRevA.100.023806} {\bibfield
   {journal} {\bibinfo  {journal} {Phys. Rev. A}\ }\textbf {\bibinfo {volume}
  {100}},\ \bibinfo {pages} {023806} (\bibinfo {year} {2019})}\BibitemShut
  {NoStop}%
\bibitem [{\citenamefont {Needham}\ \emph {et~al.}(2019)\citenamefont
  {Needham}, \citenamefont {Lesanovsky},\ and\ \citenamefont
  {Olmos}}]{Needham19}%
  \BibitemOpen
  \bibfield  {author} {\bibinfo {author} {\bibfnamefont {Jemma~A}\ \bibnamefont
  {Needham}}, \bibinfo {author} {\bibfnamefont {Igor}\ \bibnamefont
  {Lesanovsky}}, \ and\ \bibinfo {author} {\bibfnamefont {Beatriz}\
  \bibnamefont {Olmos}},\ }\bibfield  {title} {\enquote {\bibinfo {title}
  {Subradiance-protected excitation transport},}\ }\href {\doibase
  10.1088/1367-2630/ab31e8} {\bibfield  {journal} {\bibinfo  {journal} {New
  Journal of Physics}\ }\textbf {\bibinfo {volume} {21}},\ \bibinfo {pages}
  {073061} (\bibinfo {year} {2019})}\BibitemShut {NoStop}%
\bibitem [{\citenamefont {Adamo}\ \emph {et~al.}(2012)\citenamefont {Adamo},
  \citenamefont {Ou}, \citenamefont {So}, \citenamefont {Jenkins},
  \citenamefont {De~Angelis}, \citenamefont {MacDonald}, \citenamefont
  {Di~Fabrizio}, \citenamefont {Ruostekoski},\ and\ \citenamefont
  {Zheludev}}]{AdamoEtAlPRL2012}%
  \BibitemOpen
  \bibfield  {author} {\bibinfo {author} {\bibfnamefont {G.}~\bibnamefont
  {Adamo}}, \bibinfo {author} {\bibfnamefont {J.~Y.}\ \bibnamefont {Ou}},
  \bibinfo {author} {\bibfnamefont {J.~K.}\ \bibnamefont {So}}, \bibinfo
  {author} {\bibfnamefont {S.~D.}\ \bibnamefont {Jenkins}}, \bibinfo {author}
  {\bibfnamefont {F.}~\bibnamefont {De~Angelis}}, \bibinfo {author}
  {\bibfnamefont {K.~F.}\ \bibnamefont {MacDonald}}, \bibinfo {author}
  {\bibfnamefont {E.}~\bibnamefont {Di~Fabrizio}}, \bibinfo {author}
  {\bibfnamefont {J.}~\bibnamefont {Ruostekoski}}, \ and\ \bibinfo {author}
  {\bibfnamefont {N.~I.}\ \bibnamefont {Zheludev}},\ }\bibfield  {title}
  {\enquote {\bibinfo {title} {Electron-beam-driven collective-mode
  metamaterial light source},}\ }\href {\doibase
  10.1103/PhysRevLett.109.217401} {\bibfield  {journal} {\bibinfo  {journal}
  {Phys. Rev. Lett.}\ }\textbf {\bibinfo {volume} {109}},\ \bibinfo {pages}
  {217401} (\bibinfo {year} {2012})}\BibitemShut {NoStop}%
\bibitem [{\citenamefont {Lehmberg}(1970)}]{Lehmberg1970}%
  \BibitemOpen
  \bibfield  {author} {\bibinfo {author} {\bibfnamefont {R.~H.}\ \bibnamefont
  {Lehmberg}},\ }\bibfield  {title} {\enquote {\bibinfo {title} {Radiation from
  an $n$-atom system. i. general formalism},}\ }\href {\doibase
  10.1103/PhysRevA.2.883} {\bibfield  {journal} {\bibinfo  {journal} {Phys.
  Rev. A}\ }\textbf {\bibinfo {volume} {2}},\ \bibinfo {pages} {883--888}
  (\bibinfo {year} {1970})}\BibitemShut {NoStop}%
\bibitem [{\citenamefont {Jackson}(1999)}]{Jackson}%
  \BibitemOpen
  \bibfield  {author} {\bibinfo {author} {\bibfnamefont {John~David}\
  \bibnamefont {Jackson}},\ }\href@noop {} {\emph {\bibinfo {title} {Classical
  Electrodynamics}}},\ \bibinfo {edition} {3rd}\ ed.\ (\bibinfo  {publisher}
  {Wiley, New York},\ \bibinfo {year} {1999})\BibitemShut {NoStop}%
\bibitem [{\citenamefont {Lee}\ \emph {et~al.}(2016)\citenamefont {Lee},
  \citenamefont {Jenkins},\ and\ \citenamefont {Ruostekoski}}]{Lee16}%
  \BibitemOpen
  \bibfield  {author} {\bibinfo {author} {\bibfnamefont {Mark~D.}\ \bibnamefont
  {Lee}}, \bibinfo {author} {\bibfnamefont {Stewart~D.}\ \bibnamefont
  {Jenkins}}, \ and\ \bibinfo {author} {\bibfnamefont {Janne}\ \bibnamefont
  {Ruostekoski}},\ }\bibfield  {title} {\enquote {\bibinfo {title} {Stochastic
  methods for light propagation and recurrent scattering in saturated and
  nonsaturated atomic ensembles},}\ }\href {\doibase
  10.1103/PhysRevA.93.063803} {\bibfield  {journal} {\bibinfo  {journal} {Phys.
  Rev. A}\ }\textbf {\bibinfo {volume} {93}},\ \bibinfo {pages} {063803}
  (\bibinfo {year} {2016})}\BibitemShut {NoStop}%
\bibitem [{\citenamefont {Ruostekoski}\ and\ \citenamefont
  {Javanainen}(1997{\natexlab{b}})}]{Ruostekoski1997b}%
  \BibitemOpen
  \bibfield  {author} {\bibinfo {author} {\bibfnamefont {Janne}\ \bibnamefont
  {Ruostekoski}}\ and\ \bibinfo {author} {\bibfnamefont {Juha}\ \bibnamefont
  {Javanainen}},\ }\bibfield  {title} {\enquote {\bibinfo {title}
  {Lorentz-{Lorenz} shift in a {Bose}-{Einstein} condensate},}\ }\href
  {\doibase 10.1103/PhysRevA.56.2056} {\bibfield  {journal} {\bibinfo
  {journal} {Phys. Rev. A}\ }\textbf {\bibinfo {volume} {56}},\ \bibinfo
  {pages} {2056--2059} (\bibinfo {year} {1997}{\natexlab{b}})}\BibitemShut
  {NoStop}%
\bibitem [{\citenamefont {Gerbier}\ \emph {et~al.}(2006)\citenamefont
  {Gerbier}, \citenamefont {Widera}, \citenamefont {F\"olling}, \citenamefont
  {Mandel},\ and\ \citenamefont {Bloch}}]{gerbier_pra_2006}%
  \BibitemOpen
  \bibfield  {author} {\bibinfo {author} {\bibfnamefont {Fabrice}\ \bibnamefont
  {Gerbier}}, \bibinfo {author} {\bibfnamefont {Artur}\ \bibnamefont {Widera}},
  \bibinfo {author} {\bibfnamefont {Simon}\ \bibnamefont {F\"olling}}, \bibinfo
  {author} {\bibfnamefont {Olaf}\ \bibnamefont {Mandel}}, \ and\ \bibinfo
  {author} {\bibfnamefont {Immanuel}\ \bibnamefont {Bloch}},\ }\bibfield
  {title} {\enquote {\bibinfo {title} {Resonant control of spin dynamics in
  ultracold quantum gases by microwave dressing},}\ }\href {\doibase
  10.1103/PhysRevA.73.041602} {\bibfield  {journal} {\bibinfo  {journal} {Phys.
  Rev. A}\ }\textbf {\bibinfo {volume} {73}},\ \bibinfo {pages} {041602(R)}
  (\bibinfo {year} {2006})}\BibitemShut {NoStop}%
\bibitem [{\citenamefont {Svidzinsky}\ \emph {et~al.}(2010)\citenamefont
  {Svidzinsky}, \citenamefont {Chang},\ and\ \citenamefont {Scully}}]{SVI10}%
  \BibitemOpen
  \bibfield  {author} {\bibinfo {author} {\bibfnamefont {Anatoly~A.}\
  \bibnamefont {Svidzinsky}}, \bibinfo {author} {\bibfnamefont {Jun-Tao}\
  \bibnamefont {Chang}}, \ and\ \bibinfo {author} {\bibfnamefont {Marlan~O.}\
  \bibnamefont {Scully}},\ }\bibfield  {title} {\enquote {\bibinfo {title}
  {Cooperative spontaneous emission of $n$ atoms: Many-body eigenstates, the
  effect of virtual {L}amb shift processes, and analogy with radiation of $n$
  classical oscillators},}\ }\href {\doibase 10.1103/PhysRevA.81.053821}
  {\bibfield  {journal} {\bibinfo  {journal} {Phys. Rev. A}\ }\textbf {\bibinfo
  {volume} {81}},\ \bibinfo {pages} {053821} (\bibinfo {year}
  {2010})}\BibitemShut {NoStop}%
\bibitem [{\citenamefont {Petrosyan}\ and\ \citenamefont
  {M\o{}lmer}(2018)}]{Petrosyan18}%
  \BibitemOpen
  \bibfield  {author} {\bibinfo {author} {\bibfnamefont {David}\ \bibnamefont
  {Petrosyan}}\ and\ \bibinfo {author} {\bibfnamefont {Klaus}\ \bibnamefont
  {M\o{}lmer}},\ }\bibfield  {title} {\enquote {\bibinfo {title} {Deterministic
  free-space source of single photons using rydberg atoms},}\ }\href {\doibase
  10.1103/PhysRevLett.121.123605} {\bibfield  {journal} {\bibinfo  {journal}
  {Phys. Rev. Lett.}\ }\textbf {\bibinfo {volume} {121}},\ \bibinfo {pages}
  {123605} (\bibinfo {year} {2018})}\BibitemShut {NoStop}%
\bibitem [{\citenamefont {Mandel}\ \emph {et~al.}(2003)\citenamefont {Mandel},
  \citenamefont {Greiner}, \citenamefont {Widera}, \citenamefont {Rom},
  \citenamefont {H{\"a}nsch},\ and\ \citenamefont {Bloch}}]{mandel03}%
  \BibitemOpen
  \bibfield  {author} {\bibinfo {author} {\bibfnamefont {Olaf}\ \bibnamefont
  {Mandel}}, \bibinfo {author} {\bibfnamefont {Markus}\ \bibnamefont
  {Greiner}}, \bibinfo {author} {\bibfnamefont {Artur}\ \bibnamefont {Widera}},
  \bibinfo {author} {\bibfnamefont {Tim}\ \bibnamefont {Rom}}, \bibinfo
  {author} {\bibfnamefont {Theodor~W.}\ \bibnamefont {H{\"a}nsch}}, \ and\
  \bibinfo {author} {\bibfnamefont {Immanuel}\ \bibnamefont {Bloch}},\
  }\bibfield  {title} {\enquote {\bibinfo {title} {Controlled collisions for
  multi-particle entanglement of optically trapped atoms},}\ }\href {\doibase
  10.1038/nature02008} {\bibfield  {journal} {\bibinfo  {journal} {Nature}\
  }\textbf {\bibinfo {volume} {425}},\ \bibinfo {pages} {937--940} (\bibinfo
  {year} {2003})}\BibitemShut {NoStop}%
\bibitem [{\citenamefont {Jenkins}\ \emph
  {et~al.}(2016{\natexlab{b}})\citenamefont {Jenkins}, \citenamefont
  {Ruostekoski}, \citenamefont {Javanainen}, \citenamefont {Jennewein},
  \citenamefont {Bourgain}, \citenamefont {Pellegrino}, \citenamefont
  {Sortais},\ and\ \citenamefont {Browaeys}}]{Jenkins_long16}%
  \BibitemOpen
  \bibfield  {author} {\bibinfo {author} {\bibfnamefont {S.~D.}\ \bibnamefont
  {Jenkins}}, \bibinfo {author} {\bibfnamefont {J.}~\bibnamefont
  {Ruostekoski}}, \bibinfo {author} {\bibfnamefont {J.}~\bibnamefont
  {Javanainen}}, \bibinfo {author} {\bibfnamefont {S.}~\bibnamefont
  {Jennewein}}, \bibinfo {author} {\bibfnamefont {R.}~\bibnamefont {Bourgain}},
  \bibinfo {author} {\bibfnamefont {J.}~\bibnamefont {Pellegrino}}, \bibinfo
  {author} {\bibfnamefont {Y.~R.~P.}\ \bibnamefont {Sortais}}, \ and\ \bibinfo
  {author} {\bibfnamefont {A.}~\bibnamefont {Browaeys}},\ }\bibfield  {title}
  {\enquote {\bibinfo {title} {Collective resonance fluorescence in small and
  dense atom clouds: Comparison between theory and experiment},}\ }\href
  {\doibase 10.1103/PhysRevA.94.023842} {\bibfield  {journal} {\bibinfo
  {journal} {Phys. Rev. A}\ }\textbf {\bibinfo {volume} {94}},\ \bibinfo
  {pages} {023842} (\bibinfo {year} {2016}{\natexlab{b}})}\BibitemShut
  {NoStop}%
\bibitem [{\citenamefont {Fleischhauer}\ \emph {et~al.}(2005)\citenamefont
  {Fleischhauer}, \citenamefont {Imamoglu},\ and\ \citenamefont
  {Marangos}}]{FleischhauerEtAlRMP2005}%
  \BibitemOpen
  \bibfield  {author} {\bibinfo {author} {\bibfnamefont {Michael}\ \bibnamefont
  {Fleischhauer}}, \bibinfo {author} {\bibfnamefont {Atac}\ \bibnamefont
  {Imamoglu}}, \ and\ \bibinfo {author} {\bibfnamefont {Jonathan~P.}\
  \bibnamefont {Marangos}},\ }\bibfield  {title} {\enquote {\bibinfo {title}
  {Electromagnetically induced transparency: Optics in coherent media},}\
  }\href {\doibase 10.1103/RevModPhys.77.633} {\bibfield  {journal} {\bibinfo
  {journal} {Rev. Mod. Phys.}\ }\textbf {\bibinfo {volume} {77}},\ \bibinfo
  {pages} {633--673} (\bibinfo {year} {2005})}\BibitemShut {NoStop}%
\bibitem [{\citenamefont {Liu}\ \emph {et~al.}(2001)\citenamefont {Liu},
  \citenamefont {Dutton}, \citenamefont {Behroozi},\ and\ \citenamefont
  {Hau}}]{LiuEtAlNature2001}%
  \BibitemOpen
  \bibfield  {author} {\bibinfo {author} {\bibfnamefont {C.}~\bibnamefont
  {Liu}}, \bibinfo {author} {\bibfnamefont {Z.}~\bibnamefont {Dutton}},
  \bibinfo {author} {\bibfnamefont {C.~H.}\ \bibnamefont {Behroozi}}, \ and\
  \bibinfo {author} {\bibfnamefont {L.~H.}\ \bibnamefont {Hau}},\ }\bibfield
  {title} {\enquote {\bibinfo {title} {Observation of coherent optical
  information storage in and atomic medium using halted light pulses},}\ }\href
  {https://doi.org/10.1038/35054017} {\bibfield  {journal} {\bibinfo  {journal}
  {Nature}\ }\textbf {\bibinfo {volume} {409}},\ \bibinfo {pages} {490}
  (\bibinfo {year} {2001})}\BibitemShut {NoStop}%
\bibitem [{\citenamefont {Dutton}\ and\ \citenamefont
  {Vestergaard~Hau}(2004)}]{DuttonHau}%
  \BibitemOpen
  \bibfield  {author} {\bibinfo {author} {\bibfnamefont {Zachary}\ \bibnamefont
  {Dutton}}\ and\ \bibinfo {author} {\bibfnamefont {Lene}\ \bibnamefont
  {Vestergaard~Hau}},\ }\bibfield  {title} {\enquote {\bibinfo {title} {Storing
  and processing optical information with ultraslow light in {Bose-Einstein}
  condensates},}\ }\href {\doibase 10.1103/PhysRevA.70.053831} {\bibfield
  {journal} {\bibinfo  {journal} {Phys. Rev. A}\ }\textbf {\bibinfo {volume}
  {70}},\ \bibinfo {pages} {053831} (\bibinfo {year} {2004})}\BibitemShut
  {NoStop}%
\bibitem [{\citenamefont {Javanainen}\ \emph {et~al.}(1999)\citenamefont
  {Javanainen}, \citenamefont {Ruostekoski}, \citenamefont {Vestergaard},\ and\
  \citenamefont {Francis}}]{Javanainen1999a}%
  \BibitemOpen
  \bibfield  {author} {\bibinfo {author} {\bibfnamefont {Juha}\ \bibnamefont
  {Javanainen}}, \bibinfo {author} {\bibfnamefont {Janne}\ \bibnamefont
  {Ruostekoski}}, \bibinfo {author} {\bibfnamefont {Bjarne}\ \bibnamefont
  {Vestergaard}}, \ and\ \bibinfo {author} {\bibfnamefont {Matthew~R.}\
  \bibnamefont {Francis}},\ }\bibfield  {title} {\enquote {\bibinfo {title}
  {One-dimensional modeling of light propagation in dense and degenerate
  samples},}\ }\href {\doibase 10.1103/PhysRevA.59.649} {\bibfield  {journal}
  {\bibinfo  {journal} {Phys. Rev. A}\ }\textbf {\bibinfo {volume} {59}},\
  \bibinfo {pages} {649--666} (\bibinfo {year} {1999})}\BibitemShut {NoStop}%
\bibitem [{\citenamefont {Northup}\ and\ \citenamefont
  {Blatt}(2014)}]{Northup2014}%
  \BibitemOpen
  \bibfield  {author} {\bibinfo {author} {\bibfnamefont {T.~E.}\ \bibnamefont
  {Northup}}\ and\ \bibinfo {author} {\bibfnamefont {R.}~\bibnamefont
  {Blatt}},\ }\bibfield  {title} {\enquote {\bibinfo {title} {Quantum
  information transfer using photons},}\ }\href
  {https://doi.org/10.1038/nphoton.2014.53} {\bibfield  {journal} {\bibinfo
  {journal} {Nature Photonics}\ }\textbf {\bibinfo {volume} {8}},\ \bibinfo
  {pages} {356--363} (\bibinfo {year} {2014})},\ \bibinfo {note} {review
  Article}\BibitemShut {NoStop}%
\end{thebibliography}
\end{document}